\documentclass[11pt]{article}

\usepackage{bm}
\usepackage{amssymb}
\usepackage{amsmath}
\usepackage{amsthm}
\usepackage{hyperref}
\usepackage{tikz}

\usetikzlibrary{graphs, positioning, quotes, shapes.geometric}
\usepackage{geometry}

\newtheorem{thm}{Theorem}[section]
\newtheorem{lem}[thm]{Lemma}
\newtheorem{coro}[thm]{Corollary}

\newtheorem{alg}{Algorithm}[section]

\newcommand{\br}{{\mathbb R}}

\newcommand{\bn}{{\mathbb N}}
\newcommand{\bc}{{\mathbb C}}

\newcommand{\diag}{{\rm diag}}
\newcommand{\ifs}{\mbox{\rm if~}}
\newcommand{\otherwise}{\mbox{\rm otherwise~}}
\newcommand{\prf}{\par{\bf Proof. }}
\newcommand{\bbox}{\rule{2mm}{3mm}}
\newcommand{\rank}{{\rm rank}}
\newcommand{\supp}{{\rm supp}}
\newcommand{\spann}{{\rm span}}
\newcommand{\st}{{\rm s.~t.~}}
\newcommand{\tr}{{\rm tr}}
\newcommand{\bfone}{\mathbf{1}}

\newcommand{\bfu}{\mathbf{u}}
\newcommand{\bfe}{\mathbf{e}}
\newcommand{\bfh}{\mathbf{h}}
\newcommand{\bfg}{\mathbf{g}}
\newcommand{\bfv}{\mathbf{v}}
\newcommand{\bff}{\mathbf{f}}
\newcommand{\bfx}{\mathbf{x}}
\newcommand{\bfy}{\mathbf{y}}
\newcommand{\bfA}{\mathbf{A}}
\newcommand{\bfF}{\mathbf{F}}
\newcommand{\bfB}{\mathbf{B}}
\newcommand{\bfL}{\mathbf{L}}
\newcommand{\bfU}{\mathbf{U}}

\newcommand{\bfI}{\mathbf{I}}
\newcommand{\bfJ}{\mathbf{J}}
\newcommand{\bfP}{\mathbf{P}}
\newcommand{\bfQ}{\mathbf{Q}}
\newcommand{\bfW}{\mathbf{W}}
\newcommand{\bfz}{\mathbf{z}}

\title{Perfect Reconstruction Two-Channel Filter Banks on Arbitrary 
Graphs\footnote{Supported by National Natural Science Foundation of China 
(Nos. 12171488, 11771458) and 
Guangdong Province Key Laboratory of Computational Science 
at the Sun Yat-sen University (2020B1212060032).
}}

\author{Junxia You and Lihua Yang\footnote{Corresponding author}\\
School of Mathematics, Sun Yat-sen University, Guangzhou, China\\
Guangdong Province Key Laboratory of Computational Science
}
\date{\today}

\begin{document}
\maketitle

	
\begin{abstract}
This paper extends the existing theory of perfect reconstruction 
two-channel filter banks from bipartite graphs to non-bipartite graphs. By generalizing the 
concept of downsampling/upsampling we establish the frame of two-channel filter bank on arbitrary connected, undirected and weighted graphs. Then the equations for perfect reconstruction of the filter banks are presented and solved under proper conditions.
 Algorithms for designing orthogonal and biorthogonal banks are given and two typical orthogonal  two-channel filter banks are calculated. 
The locality and
approximation properties of such filter banks are discussed theoretically and experimentally.
\par
\textbf{Keywords:} Graph signal processing, wavelets, two-channel filter banks, perfect reconstruction
\end{abstract}

\section{Introduction}
Graph signal processing (GSP) is an emerging field that studies signals defined on the 
vertices of a weighted graph: i.e.\ vertices connected by edges associated with non-negative weights \cite{shuman2013emerging, ortega2018graph}. Weighted graphs provide a natural representation for data domain in many applications, such as the social networks, web information analysis, sensor networks and machine learning. The collections of samples on these graphs are termed as \textit{graph signals}. For example, a social network can be modeled as a weighted graph by viewing the individual accounts as vertices, and the relationships between them as weighted edges. Then one can analyze the information of all the accounts in this network by using GSP tools. 
Similarly, in a sensor network, the sensors and the distances between each of them constitute a graph and the recorded data on the sensors defines a signal on the graph.
In recent years, graph signal processing technology has been widely used
\cite{ortega2018graph, jablonski2017graph, yang2020new, lin2020feature}.
Graph signal processing aims at extending the 
well-developed theory and methods for analysis of signals defined in regular domains to 
those defined in irregular graph domains. There has been a lot of research in this field, 
including the Fourier transform of graph functions \cite{sandryhaila2013discrete, deri2017spectral, yang2021graph}, graph sampling and reconstruction 
\cite{narang2013localized, huang2020reconstruction, yang2021orthogonal},
approximation theory of graph functions 
\cite{pesenson2010sampling, huang2021approximation},
graph wavelets and multiscale analysis \cite{narang2012perfect, coifman2006diffusion, hammond2011wavelets, dong2017sparse, nakahira2018parseval, gavish2010multiscale}, 
and so on.

In many applications, a certain type of transform is applied to the original signal if it brings benefits in analysis in the transformed domain than in the original signal domain. And then the processing and analysis is performed on the coefficients of the transformed data.  For processing of signals defined in the regular domains, transforms such as Fourier transform, windowed Fourier transform and wavelet transform have been developed. Among them, wavelet transform is particularly widely used for processing nonstationary
signals because it catches the local information of the signal in both time and
frequency domains.
Naturally, people want to extend the theory and methods of wavelet analysis to the graph signal processing. However, due to the irregularity of graph structure, some traditional operations such as  translation and dilation are difficult to establish in the graph settings.  But people are still actively seeking ways to develop wavelet transforms on graphs. 

In \cite{crovella2003graph}, Crovella and Kolaczyk constructed a series of simple functions on each neighbourhood of every vertex so that they are compactly supported and have zero integral over the entire vertex set. They refer to these functions as graph wavelet functions. 
Coifman and Maggioni proposed the concept of diffusion wavelets and use diffusion as a smoothing and 
scaling tool to enable coarse graining and multiscale analysis in \cite{coifman2006diffusion}.
Gavish et al. \cite{gavish2010multiscale} first constructed multiscale wavelet-like orthonormal bases on hierarchical trees. They proved that function smoothness with respect to a metric induced by the tree is equivalent to approximate sparsity. Hammond et al. \cite{hammond2011wavelets} constructed wavelet transforms in the graph domain based on the spectral graph theory, and they presented a fast Chebyshev polynomial approximation algorithm to improve efficiency. In follow-up work, they also built an almost tight wavelet frame based on the polynomial filters \cite{tay2017almost}. 
In \cite{shuman2015spectrum}, Shuman et al. proposed  filters adapted to the distribution of graph Laplacian eigenvalues, leading to atoms with better discriminatory power. 
Inspired by the first-order spline filters in classical signal processing, Ekambaram et al. designed a class of critically sampled and perfect recontruction spline wavelets, and was later extended to higher-order and exponential spline filters by Kotzagiannidis and Dragotti \cite{ekambaram2015spline, kotzagiannidis2019splines}.
In \cite{narang2012perfect},
Narang and Ortega designed perfect reconstruction two-channel filter banks on bipartite graphs based on the  spectral folding phenomenon. For non-bipartite graphs, they proposed an algorithm that can decompose any graph into a series of bipartite subgraphs, thereby extending the design to arbitrary graphs. In the follow-up work \cite{narang2013compact}, they constructed a class of biorthogonal wavelet filter banks on bipartite graphs, where all filters are polynomials in the Laplacian matrix. 
\par
When a non-bipartite graph is decomposed into several bipartite subgraphs, the signal processing 
on the original graph comes down to the signal processing on every bipartite subgraph. 
A challenging topic is: can we construct perfect reconstruction two-channel filter banks on non-bipartite graphs directly? 
Inspired by \cite{narang2012perfect}, by generalizing the concepts of
downsampling and upsampling operations, we extend the construction of
perfect reconstruction two-channel filter banks proposed in \cite{narang2012perfect} to
arbitrary connected, undirected, and weighted graphs in this paper. The locality and 
approximation property of such filter banks are discussed theoretically and experimentally.
\par
The rest of the paper is organized as follows: 
Section \ref{sec:background} introduces some basic concepts 
including the graph Fourier transform, filters, downsampling and upsampling, and 
the two-channel filter banks. The related work \cite{narang2012perfect} is also introduced
briefly in this section to motivate our work, and the contribution of this paper is summarized
at the end of this section. 
In section \ref{sec:3}, the main theorem for constructing 
perfect reconstruction two-channel filter banks on arbitrary graphs is established.
The generalized downsamplers/upsamplers are constructed and
the perfect reconstruction equations for a two-channel filter bank are presented.
Algorithms for designing orthogonal and biorthogonal filter banks 
are given and two typical orthogonal filter banks are designed. 
Finally, the locality and approximation property of the 
proposed filter banks are discussed theoretically and experimentally in Section \ref{sec:4}.

\section{Preliminary}\label{sec:background}
\subsection{Notations}
We start by introducing the notations used throughout this paper. 
Vectors are denoted by lowercase boldfaced letters and matrices are denoted by uppercase boldfaced letters. The set of real numbers and the set of natural numbers are denoted as $\br$ and $\bn$ respectively. For any $N, M\in\bn$, the linear spaces of all the 
$N$-dimensional column vectors and all the matrices of order $N\times M$ are respectively
denoted by $\br^N$ and $\br^{N\times M}$. $\br^N_+$ is the set of vectors in $\br^N$ whose
components are all non-negative.
The $i$th component of a vector $\bfx$ is denoted by $x_i$ or $\bfx(i)$. The $(i,j)$-entry of matrix $\bfA$ is denoted by ${\bfA}(i,j)$ or $a_{ij}$. Let $\mathbf{1}_N$ and $\mathbf{0}_N$ represent the vectors  in $\br^N$ whose components are all $1$ and $0$ respectively. $\bfI_N$ stands for the identity matrix of order $N$. For $1\leq i\leq N$, let $\bfe_i$ be the $i$th column of $\bfI_N$. For any $x\in\br$, $[x]$ represents the largest integer not exceeding $x$. 
\par
Let $\mathcal{G}=(\mathcal{V},\mathcal{E},\bfW)$
be a connected, undirected and weighted graph with neither loops nor multiple edges,
where $\mathcal{V}=\{v_1,...,v_N\}$ is the set of vertices, $\mathcal{E}$ is the set of edges, and $\bfW\in\br^{N\times N}$ is the adjacency matrix
with its entry $w_{ij}$ the nonnegative weight of the edge between the vertices $v_i$ and $v_j$.
A graph signal $f:\mathcal{V}\to\br$ is a function defined on the vertices of the graph. Once the vertex order is fixed, the graph signal can be written as a vector 
$\bff:=(f(v_1),...,f(v_N))^{\top}\in\br^N$, where the $i$th component equals the
value of $f$ on $v_i$. In this paper, we will not
distinguish the difference between $f$ and $\bff$ if no confusion arises.
\par
\par
The superscript $^\top$ indicates the transpose operation. Function $\diag(\cdot)$ maps a 
vertor to a diagonal matrix, or a matrix to its diagonal.
We denote by $\langle {\bfv},{\bfu}\rangle$ the inner product of the vectors $\bfu$ 
and $\bfv$ in the Euclidean space $\br^N$. The induced norm is called $2$-norm and denoted by 
$\|\bfu\|_2$.
We adopt the following Dirichlet form to measure the oscillation of a graph signal $\bff$ on $\mathcal{G}$ \cite{shuman2013emerging}:
\begin{equation}\label{pDirichlet}
S_2(\bff):=\frac{1}{2}\sum^N_{i=1}\sum^N_{j=1}w_{ij}|\bff(v_i)-\bff(v_j)|^2.
\end{equation}
It is easy to see that the larger the value of $S_2(\bff)$, the stronger the signal oscillates
and vice versa.

\subsection{Fourier Transform and Filters}
\label{sec:FT}
The Laplacian matrix of a graph $\mathcal{G}=(\mathcal{V},\mathcal{E},\bfW)$
is defined as $\bfL:=\mathbf{D}-\bfW$,
where $\mathbf{D}$ is the diagonal degree matrix $\diag(d_1,...,d_N)$ with elements
$d_i=\sum_{j=1}^N w_{ij}$ \cite{chung1997spectral}. 
As the matrix ${\bfL}$ is real symmetric and positive semi-definite,
there exists a set of orthonormal eigenvectors $\{\bfu_l\}^N_{l=1}$ and real eigenvalues $0=\lambda_1<\lambda_2\le\cdots\le\lambda_N$
such that $\bfL={\bfU}{\Lambda}{\bfU}^\mathrm{\top}$, where $${\bfU}:=(\bfu_1,...,\bfu_N),~~~~
{\Lambda}:=\diag(\lambda_1,...,\lambda_N).$$
The set of the eigenvectors $\{\bfu_l\}^N_{l=1}$ are often viewed as the graph Fourier 
basis and $\bfU$ is called the Fourier basis matrix.
Using the Fourier basis, the graph Fourier transform and the inverse Fourier transform 
are defined respectively as \cite{shuman2013emerging}:
$$\hat{{\bff}}:={\bfU}^{\top}{\bff},~~~~
\bff={\bfU}\hat{{\bff}},~~~~\forall\, {\bff}\in\br^N.$$
\par
With the Laplacian matrix $\bfL$, 
the Dirichlet form \eqref{pDirichlet} can be rewritten as
\begin{equation}\label{eq:smooth-in-freq}
S_2(\bff)=\bff^\top\bfL\bff=\hat{\bff}^\top\Lambda\hat{\bff}
=\sum^N_{k=1}\lambda_k|\hat{\bff}(k)|^2.
\end{equation}
Since $S_2(\bfu_l)=\bfu_l^{\top}\bfL \bfu_l=\lambda_l,~l=1,...,N$, we have that
$S_2(\bfu_1)\le\cdots\le S_2(\bfu_N)$, which shows that the oscillation of the Fourier basis $\bfu_1, ..., 
\bfu_N$ becomes stronger as the index $l$ increases. In view of the above, the Dirichlet form 
$S_2(\bff)$ is regarded as the frequency of $\bff$. We call the set 
$\sigma({\bfL}):=\{\lambda_1<\lambda_2\leq\cdots\leq\lambda_N\}$
the spectra of ${\bfL}$.
\par
There are serval ways to define the Fourier transform of graph signals. In addition to the above-mentioned definition of using the eigenvectors of the graph Laplacian matrix $\bfL$, it can also be defined as the 
eigenvectors of the normalized Laplacian $\mathcal{L}:=\mathbf{D}^{-1/2}\bfL\mathbf{D}^{-1/2}$
or the adjacency matrix $\bfW$ \cite{narang2012perfect,sandryhaila2013discrete,sandryhaila2014big}.
From the perspective of minimizing the $\ell^1$ oscillation of signals,
we proposed a new definition of the graph Fourier basis in \cite{yang2021graph}, which is proved to have better sparsity.  In general, a graph Fourier basis 
$\{\bfu_1,...,\bfu_N\}$ is actually a family of graph signals, which constitute an
orthonormal basis of the  signal space $\br^N$. As the index $l$ increases, the oscillation of $\bfu_l$ intensifies, which can also be understood as a gradual increase in frequency in a sense.
\par
Filtering is the modulation of the Fourier transform of a signal, that is,
$$\bff\stackrel{\mbox{\scriptsize FT}}{\longrightarrow}
\hat{\bff}
\stackrel{\mbox{\scriptsize M}}{\longrightarrow}
\begin{pmatrix}
h_1\hat{\bff}_1\\
\vdots\\
h_N\hat{\bff}_N
\end{pmatrix}
\stackrel{\mbox{\scriptsize IFT}}{\longrightarrow}\bfF_h\bff,$$
or equivalently $\bfF_h=\bfU\diag(\bfh)\bfU^\top$,
where FT, IFT and M are the abbreviations of ``Fourier transform'', ``Inverse Fourier Transform'' and 
``Modulation''. The vector $\bfh:=(h_1,..., h_N)^{\top}$ used for
frequency modulation is called the filter vector.

\subsection{Downsampling and Uppersampling}
Downsampling (or subsampling) is the process of reducing the sampling rate of a signal.
In the classical signal processing, it is usually done by keeping the first sample and then every 
other $n$th sample after the first. In the well-known Mallat's decomposition algorithm in 
wavelet analysis, the signal is downsampled by $n=2$. In the graph signal processing, 
a downsampling operation can be defined by choosing a subset $\mathcal{V}_1\subset \mathcal{V}$ 
such that all samples of signal $\bff$ whose indices are not in $\mathcal{V}_1$ are 
discarded \cite{narang2012perfect}. That is,
\begin{equation}\label{eq:classical-DS}
{A}_{\mathcal{V}_1}:~ (f_1,..., f_N)^\top\mapsto (f_{i_1},..., f_{i_m})^\top,
\end{equation}
where $\mathcal{V}_1=\{v_{i_1}, ..., v_{i_m}\}$ is the downsampled subset and 
$A_{\mathcal{V}_1}$ is the corresponding downsampler. 
One can choose different subsets to define downsamplers according to 
different applications \cite{shuman2015multiscale, narang2010local, 
yang2021orthogonal, yang2021efficient}.
\par
To reconstruct the signal one needs to upsample the downsampled signal by 
inserting zeros to increase the sampling rate. This upsampler can be described as
\begin{equation}\label{eq:classical-US}
{B}_{\mathcal{V}_1}:~ (f_{i_1},..., f_{i_m})^\top
\mapsto(\tilde{f}_1,..., \tilde{f}_N)^\top,~~~~\mbox{where}~~~
\tilde{f}_j:=\begin{cases}
f_j & j\in\mathcal{V}_1=\{i_1,..., i_m\},\\
0 & \mathrm{\otherwise}.
\end{cases}
\end{equation}
Then the overall downsampling then upsampling operation can be illustrated by
$$\bff\to {A}_{\mathcal{V}_1}\bff\to {B}_{\mathcal{V}_1}{A}_{\mathcal{V}_1}\bff.$$
It is easy to verify that 
$${A}_{\mathcal{V}_1}^{\top}={B}_{\mathcal{V}_1}=[\bfe_{i_1}, \cdots, 
\bfe_{i_m}]\in\br^{N\times m}$$
and ${B}_{\mathcal{V}_1}{A}_{\mathcal{V}_1}$ is a diagonal 
matrix whose $i$th diagonal entry is $1$ if $v_i\in \mathcal{V}_1$ and $0$ 
otherwise, i.e., 
\begin{equation}\label{eq:BA=I+J}
{B}_{\mathcal{V}_1}{A}_{\mathcal{V}_1}
=\frac{1}{2}(\bfI_N+\mathbf{J}),
\end{equation}
where $\mathbf{J}$ is a diagonal matrix whose diagonal entries are given by
\begin{equation}\label{eq:J}
\mathbf{J}(k,k)=\begin{cases}
1 & v_k\in\mathcal{V}_1,\\
-1 & v_k\not\in\mathcal{V}_1,\\
\end{cases}~~~k=1,...,N.
\end{equation}
\par
In the following section, we will define the generalized downsampler and upsampler as matrices $\bfA\in\mathbb{R}^{m\times N}$ and $\bfB\in\mathbb{R}^{N\times m}$ with $m<N$.

\subsection{Two-Channel Filter Banks}
\label{sec:TCFB}
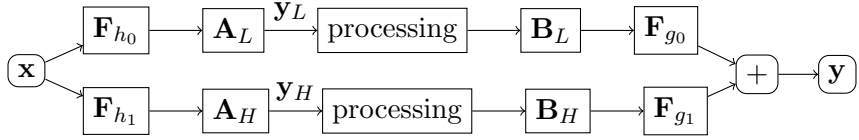
\begin{figure}[hbtp]
	\begin{center}
		\begin{tikzpicture}[node distance=10pt]
			\node[draw, rounded corners]                        (x)   {$\bfx$};
			\node[draw, above right=20pt of x,yshift=-0.5cm]         (DL)  {$\bfF_{h_0}$};
			\node[draw, right=20pt of DL]                        (SL)  {$\bfA_L$};
			\node[draw, right=20pt of SL]     					(pro1)  {processing};
			\node[draw,right=20pt of pro1]						(SLU){$\bfB_L$};
			\node[draw,right=20pt of SLU]						(RL){$\bfF_{g_0}$};
			\node[draw, below right=20pt of x,yshift=0.5cm]            (DH)  {$\bfF_{h_1}$};
			\node[draw, right=20pt of DH]                        (SH)  {$\bfA_H$};
			\node[draw, right=20pt of SH]     					(pro2)  {processing};
			\node[draw,right=20pt of pro2]						(SHU){$\bfB_H$};
			\node[draw,right=20pt of SHU]						(RH){$\bfF_{g_1}$};			
			\node[draw,rounded corners,below right=20pt of RL,yshift=0.5cm]  (plus){+};
			\node[draw, rounded corners, right=15pt of plus]  (y)     {$\bfy$};
			\graph{
				(x) ->(DL) -> (SL) -> ["$\bfy_L$"](pro1)->(SLU)->(RL) ->(plus);
				(x) ->(DH) -> (SH) -> ["$\bfy_H$"](pro2)->(SHU)->(RH) ->(plus);
				(plus)->(y);
			};
		\end{tikzpicture}
	\end{center}
\caption{A two-channel filter bank.}\label{FB}
\label{fig:2-channel-banks}
\end{figure}

A two-channel filter bank is shown in Figure  \ref{fig:2-channel-banks}. It consists of two lowpass 
filters $\bfF_{h_0}$ and $\bfF_{g_0}$, two highpass filters $\bfF_{h_1}$ and $\bfF_{g_1}$, two downsamplers
$\bfA_L, \bfA_H$ and two upsamplers $\bfB_L, \bfB_H$. 
The filters $\bfF_{h_0}$ and $\bfF_{h_1}$ are called analysis filters, and the filters $\bfF_{g_0}$ 
and $\bfF_{g_1}$ are called synthesis filters.
With a two-channel filter bank, the input signal $\bfx$ is separated into two frequency bands, a low frequency band
corresponding to the upper channel, and a high frequency band corresponding to the lower
channel. After the downsampling operation, the signal may be encoded for transmission or storage, in which case the information may be lost. 
Perfect reconstruction, i.e., $\bfy=\bfx$, requires that the analysis bank be connected directly to the synthesis bank, that is, we immediately upsample the signal after the downsampling operation \cite{winkler2000orthogonal}. A flow chart is displayed in Figure  \ref{fig:2-channel-banks}. The whole process can be mathematically expressed as \eqref{eq:TCFB}.
\begin{equation}\label{eq:TCFB}
\begin{cases}
\mbox{Input}: \bfx\\
\mbox{Analysis}: \begin{bmatrix}
			\bfy_L\\
			\bfy_H
		\end{bmatrix}:=\left[\begin{matrix}
			\bfA_L\bfF_{h_0}\\
			\bfA_H\bfF_{h_1}
		\end{matrix}\right]\bfx\\
\mbox{Reconstruction}:
\bfy:=\left[\begin{matrix}
			\bfF_{g_0}\bfB_L&\bfF_{g_1}\bfB_H		
		\end{matrix}\right]\left[\begin{matrix}
			\bfy_L\\
			\bfy_H
		\end{matrix}\right]=\bfF_{g_0}\bfB_L\bfy_L+\bfF_{g_1}\bfB_H\bfy_H\\
\mbox{Output}: \bfy=\left[\begin{matrix}
			\bfF_{g_0}\bfB_L&\bfF_{g_1}\bfB_H
		\end{matrix}\right]\left[\begin{matrix}
			\bfA_L\bfF_{h_0}\\
			\bfA_H\bfF_{h_1}
\end{matrix}\right]\bfx
\end{cases}
\end{equation}
\par
In practical applications, people need to construct different two-channel filter banks according to the application requirements. Mature mathematical theories on this subject have been developed in classical signal processing
\cite{vetterli1995wavelets}. 
In the settings of graph signal, it is still a challenging problem to design two-channel filter banks
such that the following perfect reconstruction condition is satisfied:
\begin{equation}\label{PR}
\bfF_{g_0}\bfB_L\bfA_L\bfF_{h_0}+\bfF_{g_1}\bfB_H\bfA_H\bfF_{h_1}=\bfI_N.
\end{equation}

\subsection{Related Work}\label{relatedwork}
In \cite{narang2012perfect}, Narang and Ortega established the theory of perfect reconstruction
two-channel filter banks for bipartite graph $\mathcal{G}_B=(\mathcal{V},\mathcal{E})$, where the
set of vertices $\mathcal{V}$ can be divided into two disjoint subsets $\mathcal{V}_1$
and $\mathcal{V}_2$ such that each edge in $\mathcal{E}$ connects a vertex in $\mathcal{V}_1$ to a vertex in 
$\mathcal{V}_2$.  The downsampler and upsampler in the lowpass channel are respectively 
chosen as $\bfA_L=A_{\mathcal{V}_1}$ and $\bfB_L=B_{\mathcal{V}_1}$, which are defined by
\eqref{eq:classical-DS} and \eqref{eq:classical-US}. Similarly, 
the downsampler and upsampler in the highpass channel are 
respectively chosen as $\bfA_H=A_{\mathcal{V}_2}$ and $\bfB_H=B_{\mathcal{V}_2}$. 
According to \eqref{eq:BA=I+J}, we have
$$\bfB_L\bfA_L=\frac{1}{2}(\bfI_N+\bfJ_1),~~~~
\bfB_H\bfA_H=\frac{1}{2}(\bfI_N+\bfJ_2),$$
where both $\bfJ_1$ and $\bfJ_2$ are diagonal matrix defined as follows:
\begin{equation}\label{bipart_sampling}
\mathbf{J}_i(k,k)=\begin{cases}
1 & v_k\in\mathcal{V}_i,\\
-1 & v_k\not\in\mathcal{V}_i,\end{cases}
~~k=1,...,N,~~~~i=1,2.
\end{equation}
The eigenvectors of the normalized Laplacian matrix $\mathcal{L}$ are served as
the Fourier basis in \cite{narang2012perfect}.  Let $\sigma(\mathcal{L})$ be the corresponding spectrum set. By means of the equality 
$\mathbf{J}_1+\mathbf{J}_2=\bf0$, the perfect reconstruction condition 
\eqref{PR} can be rewritten as
\begin{equation}\label{bipar_T}
\underbrace{(\bfF_{g_0}\bfF_{h_0}+\bfF_{g_1}\bfF_{h_1})}_{\mathbf{T}_1}
+\underbrace{(\bfF_{g_0}\bfJ_1\bfF_{h_0}+\bfF_{g_1}{(-\bfJ_1)\bfF}_{h_1})}_{\mathbf{T}_2}=2\bfI_N.
\end{equation}
where
\begin{align}\label{bipart_PR}
\mathbf{T}_1
&=\sum_{\lambda\in \sigma(\mathcal{L})}
(g_0(\lambda)h_0(\lambda)+g_1(\lambda)h_1(\lambda))\bfP_{\lambda},\\
\mathbf{T}_2
&=\sum_{\gamma,\lambda\in \sigma(\mathcal{L})}
(g_0(\lambda)h_0(\gamma)-g_1(\lambda)h_1(\gamma))\bfP_{\lambda}\bfJ_1\bfP_{\gamma}.
\label{eq:T2}
\end{align}
where $\bfP_{\lambda}$ is the orthogonal projector from $\br^N$ to the eigen-subspace 
$V_\lambda:=\spann\{\bfu|\mathcal{L}\bfu=\lambda \bfu\}$. 
For the bipartite graph  $\mathcal{G}_B$, since
\begin{equation}\label{eq:condition-bipartite}
\lambda\in\sigma(\mathcal{L})\iff 2-\lambda\in\sigma(\mathcal{L}),~~~~~
\bfJ_1\bfP_{\lambda}=\bfP_{2-\lambda}\bfJ_1,
\end{equation}
there holds that
$$\mathbf{T}_2
=\sum_{\lambda\in \sigma(\mathcal{L})}
\big[g_0(\lambda)h_0(2-\lambda)-g_1(\lambda)h_1(2-\lambda)\big]\bfP_{\lambda}\bfP_{2-\lambda}\bfJ_1.$$
Thus perfect reconstruction condition \eqref{bipar_T} is guaranteed by 
\begin{equation}\label{bipart_funeq}
\begin{cases}
g_0(\lambda)h_0(\lambda)+g_1(\lambda)h_1(\lambda)=2,&\\
g_1(\lambda)h_1(2-\lambda)-g_0(\lambda)h_0(2-\lambda)=0,&
\end{cases}~~\forall \lambda\in\sigma(\mathcal{L}).
\end{equation}
By setting $g_0(\lambda)=h_1(2-\lambda)$ and $g_1(\lambda)=h_0(2-\lambda)$, 
\eqref{bipart_funeq} can be simplified as
\begin{equation}\label{bipart_funeq-1}
g_0(\lambda)h_0(\lambda)+g_0(2-\lambda)h_0(2-\lambda)=2,~~\forall \lambda\in\sigma(\mathcal{L}).
\end{equation}
The resulting filter banks are said to be biorthogonal, which are studied 
in \cite{narang2013compact}. Furthermore, if
$g_i(\lambda)=h_i(\lambda)$ for $i=0,1$, then (\ref{bipart_funeq}) is equivalent to 
$$|h_0(\lambda)|^2+|h_1(\lambda)|^2=2,~~\forall \lambda\in\sigma(\mathcal{L}).$$
The filter banks satisfying the equation are said to be orthogonal, which are studied
in \cite{narang2012perfect}. The technique for the construction
is very skillful. However, it only applies to bipartite graphs due to the key condition
\eqref{eq:condition-bipartite}. For non-bipartite graphs the authors of \cite{narang2012perfect} 
proposed an approach, called Harary's decomposition, to decompose the graph into about 
$\lceil\log_2K\rceil$ bipartite subgraphs, then the filter bank is built based on each bipartite graph.
For more details, readers are  referred to \cite{narang2012perfect}.


\subsection{Our Contribution}
In this paper, we extend the spectral folding property \eqref{eq:condition-bipartite} and the perfect reconstruction condition \eqref{bipart_funeq} on bipartite graphs to arbitrary graphs. Specifically, for a given graph Fourier basis, we designed an orthogonal matrix $\bfQ$ which plays the role of the above matrix $\bfJ$, such that $\bfQ$ and the projection matrix $\bfP_{\lambda}$ satisfy a commutative equation similar to \eqref{eq:condition-bipartite}. Furthermore, we construct the generalized up/down-samplers based on $\bfQ$ and propose the perfect reconstruction equations for the two-channel filter banks on arbitrary graphs. Under proper assumption, the general solutions of the equations are given.
Theories on the approximation property and the locality of the filter banks are established.
Finally, experiments for two special two-channel filter banks are conducted to verify the theoretical results.

\section{Two-channel Filter Banks for Arbitrary Graphs
\setcounter{equation}{0}
\label{sec:3}}

\subsection{Two-channel Filter Banks Based on Generalized Samplers}
\label{sec:main_theory}
In this section, we will construct the perfect reconstruction two-channel filter banks for arbitrary connected, weighted and 
undirected graphs. As discussed above, the condition \eqref{eq:condition-bipartite} is generally no longer
valid, which makes it difficult to derive the perfect reconstruction condition \eqref{bipart_funeq}
by requiring the matrix $\mathbf{T}_2$ defined in \eqref{eq:T2} to be zero. In order to overcome this inherent obstacle of 
non-bipartite graphs, we generalize the downsampler and upsampler
defined by \eqref{eq:classical-DS} and \eqref{eq:classical-US} to a pair of matrices
$\bfA\in\br^{m\times N}$ and $\bfB\in\br^{N\times m}$ with $m\approx N/2$, and study the 
perfect reconstruction condition \eqref{PR}. That is, we hope to find proper
$\bfA_L, \bfA_H,~\bfB_L,\bfB_H$ and 
construct filters $\bfF_{h_0}, \bfF_{g_0},\bfF_{h_1}, \bfF_{g_1}$ such that the following perfect reconstruction condition holds:
\begin{equation}\label{eq:perfect-construction}
\bfF_{g_0}\bfB_L\bfA_L\bfF_{h_0}
+\bfF_{g_1}\bfB_H\bfA_H\bfF_{h_1}=\bfI_N.
\end{equation}
Inspired by the equality \eqref{eq:BA=I+J} in the case of bipartite graphs, we assume that
the downsamplers $\bfA_L, \bfA_H$ and the updsamplers $\bfB_L,\bfB_H$ meet the following conditions:
\begin{equation}\label{eq:BA-by-Q}
{\bfB_L\bfA_L}=\frac{1}{2}({\bfI_N+\bfQ}),~~~~{\bfB_H\bfA_H}=\frac{1}{2}({\bfI_N-\bfQ}),
\end{equation}
where $\bfQ$ is an orthogonal matrix to be determined. 
\par
For the sake of clearness of description, we introduce the following notations:
A partition of the Fourier basis $\{\bfu_1,...,\bfu_N\}$ is a family of disjoint subsets
$\{\mathcal{U}_\omega\}_{\omega\in\Omega}$ satisfying 
$$\bigcup_{\omega\in\Omega}\mathcal{U}_\omega=\{\bfu_1,...,\bfu_N\}.$$
With this partition, any function $h:\Omega\to\br$ determines a filter vector $\bfh
=(h_1,...,h_N)^\top$:
$$h_i:=h(\omega),~~~\bfu_i\in\mathcal{U}_\omega,~~~i=1,...,N.$$
Thus the corresponding filter $\bfF_h$ can be expressed as
$$\bfF_h=\sum^N_{i=1}h_i\bfu_i\bfu_i^\top
=\sum_{\omega\in\Omega}h(\omega)
\sum_{\bfu\in\mathcal{U}_\omega}\bfu\bfu^\top
=\sum_{\omega\in\Omega}h(\omega)\bfP_\omega,$$
where $\bfP_\omega:=\sum_{\bfu\in\mathcal{U}_\omega}\bfu\bfu^\top$ is the 
orthogonal projector from $\br^N$ to 
$\mathcal{X}_\omega:=\spann(\mathcal{U}_\omega)$. For this reason, any fucntion
$h: \Omega\to\br$ is called a filter function associated with the partition.
\begin{thm}\label{th:PR}
Let $\{\mathcal{U}_\omega\}_{\omega\in\Omega}$ be a partition of the Fourier basis 
$\{\bfu_1,...,\bfu_N\}$ and $\bfQ$ be an orthogonal matrix of order $N$ satisfying
\begin{equation}\label{eq:th:PR-cond1}
\bfQ\mathcal{X}_\omega
=\mathcal{X}_{\kappa(\omega)}~~\mbox{for}~~
\mathcal{X}_\omega:=\spann(\mathcal{U}_\omega), ~~~\forall\omega\in\Omega,
\end{equation}
where $\kappa:\Omega\to\Omega$ is a bijection.
Assume that $\bfA_L, \bfA_H, \bfB_L, \bfB_H$ are respectively
downsamplers and upsamplers satisfying \eqref{eq:BA-by-Q}.
Then \eqref{eq:perfect-construction}
holds if the filter functions $h_0,h_1, g_0, g_1$  associated with the
partition satisfy
\begin{equation}\label{eq:th:PR-cond2}
\begin{cases}
g_0(\omega)h_0(\omega)+g_1(\omega)h_1(\omega)=2,\\
g_0(\kappa(\omega))h_0(\omega)=g_1(\kappa(\omega))h_1(\omega),
\end{cases}
~~~~\forall\omega\in\Omega.
\end{equation}
\end{thm}
\prf
By inserting \eqref{eq:BA-by-Q} into \eqref{eq:perfect-construction}, the 
perfect reconstruction condition can be rewritten as
\begin{equation}\label{eq:ch2-PR-tmp1}
\bfF_{g_0}\bfF_{h_0}+\bfF_{g_1}\bfF_{h_1}+\bfF_{g_0}\bfQ\bfF_{h_0}-\bfF_{g_1}\bfQ\bfF_{h_1}=2\bfI_N.
\end{equation}
\par
For any $\omega\in\Omega$, let $\{\bfu_{i_1},...,\bfu_{i_k}\}$ 
be an orthonormal basis of $\mathcal{X}_\omega$. Then 
$\{\bfQ\bfu_{i_1},..., \bfQ\bfu_{i_k}\}$ is an orthonormal 
basis of $\mathcal{X}_{\kappa(\omega)}$. It is followed that the orthonogal projectors
$\bfP_{\mathcal{X}_\omega}:\bc^N\rightarrow \mathcal{X}_\omega$ and
$\bfP_{\mathcal{X}_{\kappa(\omega)}}:\bc^N\rightarrow \mathcal{X}_{\kappa(\omega)}$ 
can be respectively written as
$$\bfP_{\mathcal{X}_\omega}=\sum_{j=1}^{k}\bfu_{i_j}\bfu_{i_j}^{\top},~~~
\bfP_{\mathcal{X}_{\kappa(\omega)}}
=\sum_{j=1}^{k}({\bfQ\bfu}_{i_j})({\bfQ\bfu}_{i_j})^{\top},$$
which implies that
$\bfP_{\mathcal{X}_{\kappa(\omega)}}={\bfQ \bfP}_{\mathcal{X}_\omega}\bfQ^{\top}$. 
For any filter function associated with the partition: $h: \Omega\to\br$, we have
$$\bfF_{h\circ \kappa}
=\sum_{\omega\in\Omega}h(\kappa(\omega))\bfP_{\mathcal{X}_\omega}
=\bfQ^\top\Big(\sum_{\omega\in\Omega}h(\kappa(\omega))
\bfP_{\mathcal{X}_{\kappa(\omega)}}\Big)\bfQ
=\bfQ^\top\Big(\sum_{\omega\in\Omega}h(\omega)
\bfP_{\mathcal{X}_{\omega}}\Big)\bfQ
=\bfQ^\top\bfF_{h}\bfQ,
$$
i.e., $\bfQ\bfF_{h\circ \kappa}=\bfF_{h}\bfQ$. 
Consequently there holds
$$\bfF_{g_0}\bfQ\bfF_{h_0}-\bfF_{g_1}\bfQ\bfF_{h_1}
=\bfQ\big(\bfF_{g_0\circ\kappa}\bfF_{h_0}-\bfF_{g_1\circ\kappa}\bfF_{h_1}\big).$$
Thus, \eqref{eq:ch2-PR-tmp1} is equivalent to
\begin{equation}\label{eq:ch2-PR-tmp2}
\bfF_{g_0}\bfF_{h_0}+\bfF_{g_1}\bfF_{h_1}+
\bfQ\big(\bfF_{g_0\circ\kappa}\bfF_{h_0}-\bfF_{g_1\circ\kappa}\bfF_{h_1}\big)
=2\bfI_N.
\end{equation}
Since 
$\bfP_\omega\bfP_{\omega'}=\delta_{\omega,\omega'}\bfP_\omega$, 
where $\delta_{\omega,\omega'}$ is the  Kronecker delta function, we have
$$\bfF_{g_0}\bfF_{h_0}
=\sum_{\omega\in\Omega}\sum_{\omega'\in\Omega}
g_0(\omega)h_0(\omega')\bfP_\omega\bfP_{\omega'}
=\sum_{\omega\in\Omega}g_0(\omega)h_0(\omega)\bfP_\omega.$$
Similar results hold for $\bfF_{g_1}\bfF_{h_1}, 
\bfF_{g_0\circ\kappa}\bfF_{h_0}$ and $\bfF_{g_1\circ\kappa}\bfF_{h_1}$. 
Using $\sum_{\omega\in\Omega}\bfP_\omega=\bfI_N$, we conclude that
\eqref{eq:ch2-PR-tmp2} can be rewritten as
$$\sum_{\omega\in\Omega}
\big[g_0(\omega)h_0(\omega)+g_1(\omega)h_1(\omega)-2\big]\bfP_\omega
+\bfQ\sum_{\omega\in\Omega}
\big[g_0(\kappa(\omega))h_0(\omega)-g_1(\kappa(\omega))h_1(\omega)
\big]\bfP_\omega=0.$$
This equality is guaranteed by \eqref{eq:th:PR-cond2}, obviously. The proof
is complete.
\bbox
\par
If the graph is bipartite and the Fourier basis $\bfu_1,..., \bfu_N$ are
the eigenvectors of the normalized  Laplacian matrix $\mathcal{L}$ then 
the conditions of Theorem \ref{th:PR} are satisfied. In fact, by computing the entries we 
can verify that the adjacency matrix $\bfW$ satisfies $\mathbf{WJ}_1+\bfJ_1\bfW=\bf0$
for $\bfJ_1$ defined by \eqref{bipart_sampling}, which together with 
$\bfJ_1\mathbf{D}=\mathbf{D}\bfJ_1$ implies that 
$\mathcal{L}\bfJ_1=2\bfJ_1-\bfJ_1\mathcal{L}$.
Hence, $\mathcal{L}\bfu=\lambda\bfu$ if and only if 
$\mathcal{L}\bfJ_1\bfu=(2-\lambda)\bfJ_1\bfu$.
That means, for any $\lambda\in\sigma(\mathcal{L})$,
$$\bfJ_1\mathcal{X}_\lambda=\mathcal{X}_{\kappa(\lambda)},~~~~
\mbox{with}~~\kappa:\sigma(\mathcal{L})\to\sigma(\mathcal{L}),
~~\kappa(\lambda):=2-\lambda,$$
where $\sigma(\mathcal{L})$ is the spectra of $\mathcal{L}$
and $\mathcal{X}_\lambda$ is the eigen-space associated with $\lambda$. Furthermore, according to the definitions of $\{B_{\mathcal{V}_i},A_{\mathcal{V}_i}\}_{i=1,2}$, we have
$$B_{\mathcal{V}_1}A_{\mathcal{V}_1}
=\frac{1}{2}(\bfI_N+\bfJ_1),~~~~
B_{\mathcal{V}_2}A_{\mathcal{V}_2}
=\frac{1}{2}(\bfI_N-\bfJ_1).$$
It is easy to verify that the conditions of Theorem \ref{th:PR} are satisfied 
for $\bfQ=:\bfJ_1,~\Omega:=\sigma(\mathcal{L}),~\kappa(\omega)=2-\omega$,
the downsamplers $A_{\mathcal{V}_1}, A_{\mathcal{V}_2}$ defined by \eqref{eq:classical-DS} and
the uppersamplers $B_{\mathcal{V}_1},~B_{\mathcal{V}_2}$ defined by
\eqref{eq:classical-US}.
By Theorem \ref{th:PR}, we obtain the perfect reconstruction
condition \eqref{eq:th:PR-cond2}, which is exactly the \eqref{bipart_funeq}
presented in \cite{narang2012perfect}.
\par
\begin{figure}[hbtp]
\centerline{
\includegraphics[scale=0.25]{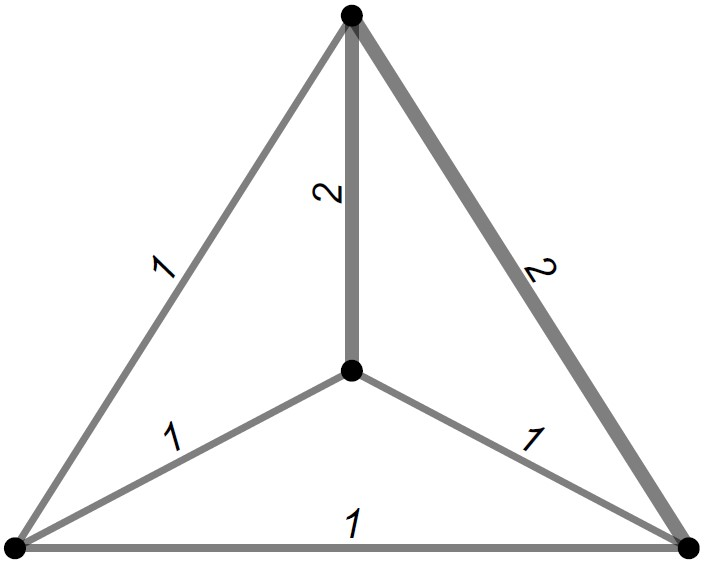}}
\caption{\small A graph of four vertices}
\label{fig:graph-4vertices}
\end{figure}

The orthogonal matrix $\bfQ$ satisfying $\bfQ\mathcal{X}_\lambda=\mathcal{X}_{\kappa(\lambda)}$
can be chosen as a diagonal matrix $\bfJ_1$ for bipartite graphs.
This fact no longer holds for non-bipartite graphs, no matter whether the Fourier transform is defined by
the normalized or non-normalized Laplacian matrix, unless $\bfQ=\pm \bfI_N$. 
We give an example to illustrate this. Let
us consider a graph with $4$ vertices as shown in Figure  \ref{fig:graph-4vertices}, whose Laplacian
matrix is given by 
$$\bfL=\begin{bmatrix}
			4  &  -1  &  -1 &   -2\\
			-1 &    3 &   -1 &   -1\\
			-1 &   -1 &    4 &   -2\\
			-2 &   -1 &   -2 &    5
\end{bmatrix}.$$
Then the eigendecomposition $\bfL={\bfU\Lambda \bfU}^{\top}$ gives the following eigenvectors 
and eigenvalues:
$$\bfU=\begin{bmatrix}
			-0.5000 &   0.2887 &   0.7071 &   0.4082\\
			-0.5000 &  -0.8660 &   0.0000 &  -0.0000\\
			-0.5000 &   0.2887 &  -0.7071 &   0.4082\\
			-0.5000 &   0.2887 &        0 &  -0.8165
\end{bmatrix},~~~
{\Lambda}=\begin{bmatrix}
			0  & 0  & 0 & 0 \\
			0  & 4  & 0 & 0 \\
			0  & 0  & 5 & 0 \\
			0  & 0  & 0 & 7 \\
\end{bmatrix}.$$
Since the four numbers of each row of ${\bfU}$ have different absolute values,
there does not exist diagonal matrix $\bfQ$ with diagonal entries $\pm 1$
such that $\{\bfQ{\bfu}_i\}_{i=1}^4$ are still eigenvectors of ${\bfL}$. 
Similar result holds for normalized Laplacian, which shows that there is no diagonal 
matrix $\bfQ$ with diagonal entries $\pm 1$ that satisfies $\bfQ\mathcal{X}_\lambda=\mathcal{X}_{\kappa(\lambda)}$.
\par
For any $\omega\in\Omega$, let $\bfU_\omega$ be the submatrix of $\bfU$
whose columns constitute a basis of $\mathcal{X}_{\omega}$. Then 
the condition \eqref{eq:th:PR-cond1} is equivalent to the existence of orthogonal matrices 
$\{A_\omega\}_{\omega\in\Omega}$ such that
$$\bfQ\bfU_\omega=\bfU_{\kappa(\omega)} A_\omega,~~~\omega\in\Omega.$$
Let $\Omega:=\{\omega_1,...,\omega_n\}$. Then
$$\bfQ[\bfU_{\omega_1},...,\bfU_{\omega_n}]
=[\bfQ\bfU_{\omega_1},...,\bfQ\bfU_{\omega_n}]
=[\bfU_{\kappa(\omega_1)},...,\bfU_{\kappa(\omega_n)}]A_\Omega~~
\mbox{with}~~A_\Omega:=
\begin{bmatrix}
A_{\omega_1}\\
& \ddots\\
&&A_{\omega_n}
\end{bmatrix}.$$
Let $\Phi$ be a block permutation and $\bfP$ be a permutation satisfying
$$[\bfU_{\kappa(\omega_1)},...,\bfU_{\kappa(\omega_n)}]=
[\bfU_{\omega_1},...,\bfU_{\omega_n}]\Phi,~~~~
[\bfU_{\omega_1},...,\bfU_{\omega_n}]=\bfU\bfP.$$
Then
\begin{equation}\label{eq:th:PR-cond3}
\bfQ\bfU
=\bfU\bfP\Phi A_\Omega\bfP^\top.
\end{equation}
\par
As a special case, let us consider the following partition of the Fourier basis:
$$\mathcal{U}_i=\spann\{\bfu_i\},~~~~i\in\Omega:=\{1,...,N\},$$
and the orthogonal matrices $\{A_{\omega_k}\}$ are positive definite. 
In this case, we have that $A_\Omega=\bfI_N$ and the condition \eqref{eq:th:PR-cond3} can 
be rewritten as $\bfQ\bfU=\bfU\Phi$, where $\Phi$ is a permutation matrix of order $N$. By
Theorem \ref{th:PR}, we have the following corollary.
\begin{coro}\label{coro:PR-a}
Let $\bfU$ be a Fourier basis matrix and $\bfQ$ an orthogonal matrix of 
order $N$ satisfying
\begin{equation}\label{eq:th:PR-cond1-a}
\bfQ\bfU=\bfU\Phi
\end{equation}
for a permutation matrix $\Phi$. Assume $\bfA_L, \bfA_H, \bfB_L, \bfB_H$ are respectively
downsamplers and upsamplers satisfying \eqref{eq:BA-by-Q}.
Then the perfect reconstruction condition \eqref{eq:perfect-construction}
holds if 
\begin{equation}\label{eq:th:PR-cond2-a}
\bfg_0\odot \bfh_0+\bfg_1\odot \bfh_1=2\bfone_N,~~~~
(\Phi^\top\bfg_0)\odot\bfh_0=(\Phi^\top\bfg_1)\odot\bfh_1,
\end{equation}
where $\odot$ stands for the Hadamard product.
\end{coro}
\par
Hereafter, unless otherwise noted, we will use the eigenvectors of the non-normalized Laplacian matrix as the Fourier basis.

\subsection{Construction of Two-channel Filter Banks for Arbitrary Graphs}
\label{sec:construction}

\subsubsection{Construction of $\bfQ$}
\label{sec:construction-Q}
In this section, we will use Corollary \ref{coro:PR-a} to construct a perfect reconstruction two-channel filter bank. To do this, we need to 
construct an orthogonal matrix $\bfQ$ satisfying \eqref{eq:th:PR-cond1-a} and proper
downsamplers $\bfA_L, \bfA_H$ and upsamplers $\bfB_L, \bfB_H$ with sizes
$\bfA_L, \bfB_L^\top\in\br^{m\times N}$ and $\bfA_H, \bfB_H^\top\in\br^{(N-m)\times N}$
for $m\approx N/2$ such that \eqref{eq:BA-by-Q} holds. 
Generally, the downsamplers $\bfA_L, \bfA_H$ are supposed
to be full row rank and the upsamplers $\bfB_L, \bfB_H$ 
to be full column rank. According to \eqref{eq:BA-by-Q}, the ranks of 
$\bfI+\bfQ$ and $\bfI-\bfQ$ should be approximately equal to $N/2$. Since
$\bfI\pm\bfQ=\bfU(\bfI\pm\Phi)\bfU^\top$, the problem turns into finding
a permutation matrix $\Phi$ such that $\rank(\bfI\pm\Phi)\approx N/2$.
\par
In the construction of the samplers in Section \ref{sec:construction-sampling},
the eigenvalues of the matrix $\Phi$ are required to
be real. For this reason, the matrix $\Phi$ is always assumed to be symmetric in the rest of this paper.
\begin{lem}\label{lem:I+Phi}
Let $\Phi$ be a symmetric permutation matrix of order $N$. Then its eigenvalues are $1$ or $-1$,
and the ranks of $\bfI\pm\Phi$ and the trace of $\Phi$ are given by
$$\rank(\bfI+\Phi)=m,~~~~\rank(\bfI-\Phi)=N-m,~~~~\tr(\Phi)=2m-N,$$
where $m$ is the algebraic multiplicity of eigenvalue $1$ of $\Phi$.
\end{lem}
\prf
Let $\lambda$ be an eigenvalue of $\Phi$, then there is a unit vector $\bfx$ such that 
$\Phi \bfx=\lambda \bfx$. Calculate the $2$-norms of the vectors on both sides we 
get $|\lambda|=\|\Phi x\|_2=\|x\|_2=1$, which yields $\lambda=\pm 1$. 
Therefore, $\Phi$ has the following Jordan decomposition:
$$\Phi=\bfP\begin{bmatrix}
\bfJ_1\\
& \bfJ_{-1}
\end{bmatrix}\bfP^{-1},$$
where $\bfP$ is an invertible matrix, $\bfJ_1, \bfJ_{-1}$ are the Jordan matrices associated 
to the eigenvalues $1$ and $-1$. Let $m$ be the order of $\bfJ_1$. Since
$$\bfI+\Phi=\bfP\begin{bmatrix}
\bfI_m+\bfJ_1\\
& \bfI_{N-m}+\bfJ_{-1}
\end{bmatrix}\bfP^{-1},~~
\bfI-\Phi=\bfP\begin{bmatrix}
\bfI_m-\bfJ_1\\
& \bfI_{N-m}-\bfJ_{-1}
\end{bmatrix}\bfP^{-1},$$
we have $\rank(\bfI+\Phi)=m$ and $\rank(\bfI-\Phi)=N-m$.
\par
Finally, it is easy to see that $\tr(\Phi)=m-(N-m)=2m-N$.
\bbox
\par
According to Lemma \ref{lem:I+Phi}, we need to construct a symmetric permutation matrix $\Phi$
whose eigenvalue $1$ has algebraic multiplicity $m\approx N/2$. That means the trace of $\Phi$ 
should be close to $0$. It is easy to see that the following matrix 
\begin{equation}\label{eq:Phi}
\Phi:=[\bfe_N,...,\bfe_1]=\begin{bmatrix}
& & 1\\
&\rotatebox[]{45}{$\cdots$}&\\
1
\end{bmatrix}\in\br^{N\times N}
\end{equation}
satisfies our requirement since its trace is either $0$ or $1$. In this case,
the algebraic multiplicity of $1$ and $-1$ are respectively 
$[(N+1)/2]$ and $[N/2]$, both are  approximately $N/2$.

\subsubsection{Construction of Generalized Sampling Matrices}
\label{sec:construction-sampling}
As discussed in Section \ref{sec:construction-Q}, $\bfQ$ can be chosen as $\bfU\Phi \bfU^{\top}$ 
where $\Phi$ is defined by \eqref{eq:Phi}. It is easy to see that
\eqref{eq:th:PR-cond1-a} holds. According to Corollary \ref{coro:PR-a}, as long 
as $\bfQ$ satisfies \eqref{eq:BA-by-Q}, i.e.,
$${\bfB_L\bfA_L}=\frac{1}{2}({\bfI_N+\bfQ}),~~~~
{\bfB_H\bfA_H}=\frac{1}{2}({\bfI_N-\bfQ})$$
with downsamplers $\bfA_L, \bfA_H$ and upsamplers $\bfB_L, \bfB_H$, 
the perfect reconstruction two-channel filter bank can be obtained by solving
the filter equations \eqref{eq:th:PR-cond2-a}. In the rest of this section, we focus on
the construction of  $\bfA_L, \bfA_H$ and $\bfB_L, \bfB_H$ that satisfy
\eqref{eq:BA-by-Q}.
\par
Let
$$\begin{cases}
r:=\big[\frac{N}{2}\big],\\
s:=\big[\frac{N+1}{2}\big],
\end{cases}~~~~
\Phi_r:=\begin{bmatrix}
& & 1\\
&\rotatebox[]{45}{$\cdots$}&\\
1
\end{bmatrix}\in\br^{r\times r}.$$
It is easy to see that $r+s=N$. 
\par
(1) If $N$ is an even number, then
$$\bfI_N+{\Phi}
=\begin{bmatrix}
\bfI_r & \Phi_r\\
\Phi_r  & \bfI_r\end{bmatrix}
=\bfP_0\bfP_0^{\top},~~~~
\bfI_N-{\Phi}
=\begin{bmatrix}
\bfI_r & -\Phi_r\\
-\Phi_r &  \bfI_r
\end{bmatrix}
=\bfP_1\bfP_1^{\top},$$
where
\begin{equation}\label{eq:P0-P1-def-even}
\bfP_0:=\begin{bmatrix}
\bfI_r \\
\Phi_r
\end{bmatrix},~~~~
\bfP_1:=\begin{bmatrix}
\bfI_r \\
-\Phi_r
\end{bmatrix}
\end{equation}
\par
(2) If $N$ is an odd number, similarly we have 
$$\bfI_N+{\Phi}
=\begin{bmatrix}
\bfI_r & \bf0 & \Phi_r\\
\bf0 & 2 & \bf0\\
\Phi_r & \bf0 & \bfI_r\end{bmatrix}
=\bfP_0\bfP_0^{\top},~~~~
\bfI_N-{\Phi}
=\begin{bmatrix}
\bfI_r & \bf0 & -\Phi_r\\
\bf0 & 0 & \bf0\\
-\Phi_r & \bf0 & \bfI_r
\end{bmatrix}
=\bfP_1\bfP_1^{\top},
$$
where
\begin{equation}\label{eq:P0-P1-def-odd}
\bfP_0:=\begin{bmatrix}
\bfI_r & \bf0\\
\bf0 & \sqrt{2}\\
\Phi_r & \bf0\end{bmatrix},~~~~
\bfP_1:=\begin{bmatrix}
\bfI_r \\
\bf0 \\
-\Phi_r
\end{bmatrix}.
\end{equation}
\par
In summary, no matter whether $N$ is even or odd, there always exist matrices 
$\bfP_0\in\br^{N\times s}$ and $\bfP_1\in\br^{N\times r}$ such that
$$\bfI_N+{\Phi}={\bfP}_0{\bfP}_0^{\top},~~~
\bfI_N-{\Phi}={\bfP}_1{\bfP}_1^{\top}.$$
which leads to 
\begin{align*}
\frac{1}{2}(\bfI_N+\bfQ)=(\frac{1}{\sqrt{2}}\mathbf{UP}_0)(\frac{1}{\sqrt{2}}\mathbf{UP}_0)^{\top}
=:\bfB_L\bfA_L,\\
\frac{1}{2}(\bfI_N-\bfQ)=(\frac{1}{\sqrt{2}}\mathbf{UP}_1)(\frac{1}{\sqrt{2}}\mathbf{UP}_1)^{\top}
=:\bfB_H\bfA_H.
\end{align*}
where
\begin{equation}\label{eq:ALAH}
\bfA_L=\bfB_L^{\top}=\frac{1}{\sqrt{2}}\bfU_1\bfP_0^\top\bfU^\top,~~~
\bfA_H=\bfB_H^{\top}=\frac{1}{\sqrt{2}}\bfP_1^\top\bfU^\top
\end{equation}
and $\bfU_1$ is an orthogonal matrix of order $s$, which will be explained and determined in the 
next section.

\subsubsection{Graph Reduction}
\begin{figure}[hbtp]
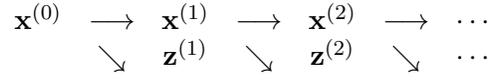

$$\begin{array}{ccccccccc}
\bfx^{(0)} &\longrightarrow & \bfx^{(1)}&
        \longrightarrow &\bfx^{(2)}&
        \longrightarrow &\cdots \\
       &\searrow & &\searrow & &\searrow  \\[-5mm]
       & & \mathbf{z}^{(1)} & & \mathbf{z}^{(2)} & &\cdots 
       \end{array} $$
\caption{Mallat's decomposition}       
\label{fig:Mallat-algorithm}       
\end{figure}

In the classical two-channel subband filtering scheme, an incoming signal $\bfx^{(0)}$
is convolved with a lowpass filter $\bfh_0$ and a highpass filter $\bfh_1$, respectively. 
Then the two resulting signals are downsampled by taking the samples in turn to produce two signals $\bfx^{(1)}$ and $\bfz^{(1)}$ of half size of $\bfx^{(0)}$. They are respectively viewed 
as a coarser approximation and a difference between $\bfx^{(0)}$ and $\bfx^{(1)}$ since 
the filter $\bfh_0$ removes the high frequency components of $\bfx^{(0)}$ while $\bfh_1$ preserves the high frequency components. The coarser approximation $\bfx^{(1)}$, as a short one-dimensional signal,
can be further repeatedly decomposed to produce coarser approximations, as illustrated by
Figure  \ref{fig:Mallat-algorithm} .
\par
As described in Section \ref{sec:TCFB}, through a two-channel filter bank, 
a graph signal $\bfx$ can be decomposed into two shorter vectors: a coarse approximation $\bfy_L$ 
and a details part $\bfy_H$ which contains the information about the difference between $\bfx$ and $\bfy_L$. 
To further decompose $\bfy_L$ into a coarser approximation of $\bfx$ in the next level, we need to equip $\bfy_L$  with a reduced graph that has a similar adjacency relationship to the original graph. This process of constructing a reduced graph is called graph reduction.
\par
There are mainly two types of graph reduction. One is to select a subset of vertices of the original 
graph followed by re-wiring. The other is to aggregate some vertices into a new vertex followed by re-wiring. There are some works about graph reduction such as \cite{sanders2011engineering}, \cite{safro2015advanced, dorfler2012kron, loukas2019graph}, \cite{ron2011relaxation}. In this paper, we use the graph coarsening method proposed by \cite{loukas2019graph}.  Given a graph Laplaican $\bfL\in\br^{N\times N}$ and a
number $s\approx N/2$, the Laplacian $\bfL_1$ of a graph $\mathcal{G}_1$ with $s$ vertices and 
similar structure to $\mathcal{G}$ can be constructed. 
Suppose the eigendecomposition of $\bfL_1$ is 
$${\bfL}_1={\bfU}_1\Lambda_1{\bfU}_1^{\top}.$$
Then, with $\bfU_1$ the samplers $\bfA_L$ and $\bfB_L$ are determined by \eqref{eq:ALAH}.

\subsubsection{Construction of Filters}
\label{cons_filter}
Let us turn to the filter equation \eqref{eq:th:PR-cond2-a}, i.e.
$$\begin{cases}
\bfg_0(k)\bfh_0(k)+\bfg_1(k)\bfh_1(k)=2,\\
\bfg_0(N+1-k)\bfh_0(k)=\bfg_1(N+1-k)\bfh_1(k),
\end{cases}~~~~k=1,...,N.$$
For simplicity, we consider the following special filter bank:
$$\bfg_0(k)=\bfh_1(N+1-k),~~~\bfg_1(k)=\bfh_0(N+1-k),~~~~k=1, ..., N.$$
Under these assumptions, Equation \eqref{eq:th:PR-cond2-a}
is equivalent to 
\begin{equation}\label{perfect-reconstruction}
\bfh_0(k)\bfg_0(k)+\bfh_0(N+1-k)\bfg_0(N+1-k)=2,~~~~k=1, ..., N.
\end{equation}
It is interesting to note that, Equation \eqref{perfect-reconstruction} 
looks like the perfect reconstruction equation of the classical biorthogonal wavelet bases
\cite{daubechies1992ten}:
$$\overline{m_0(\xi)}\tilde{m}_0(\xi)+\overline{m_0(\xi+\pi)}\tilde{m}_0(\xi+\pi)=1.$$
Thus, we refer to a filter bank $\{\bfh_0, \bfg_0, \bfh_1, \bfg_1\}$ satisfying \eqref{perfect-reconstruction} as a biorthognal filter bank.
Let $\bff(k):=\bfh_0(k)\bfg_0(k)$. Then \eqref{perfect-reconstruction}
can be rewritten as
\begin{equation}\label{perfect-reconstruction-f}
\bff(k)+\bff(N+1-k)=2,~~~~k=1, ..., N.
\end{equation}
The general solution of \eqref{perfect-reconstruction-f} is:
$$\bff(N+1-k)=2-\bff(k),~~~~k=1,...,s,$$
where $\bff(1),..., \bff(r)$ are free variables and $\bff(s)=1$ if
$N$ is an odd number.
With $\bff$, the vectors $\bfh_0$ and $\bfg_0$ can be solved from $\bfh_0\odot \bfg_0=\bff$, i.e.,
$$\bfh_0(k)\bfg_0(k)=\bff(k),~~~~k=1,...,N.$$
\par
Particularly, if the analysis filters and the synthesis filters are the same, that is,
$$\bfh_0(k)=\bfg_0(k)=\sqrt{\bff(k)},~~~~k=1,...,N,$$
then the filter bank is said to be orthogonal. In this case, since $\bff(k)=|\bfh_0(k)|^2\ge 0$, we 
have
$$0\le \bff(k)\le 2,~~~~k=1,...,s.$$
\par
In Section \ref{sec:locality} we will talk about the locality of the filter, where a filter vector 
$\bfh$ is
desired to be expressed as or approximated by a polynomial in $\lambda\in\sigma(\bfL)$. For this
purpose, $\bfh$ is assumed to satisfy
$$\lambda_i=\lambda_j\implies h_i=h_j~~\mbox{and}~~h_{N+1-i}=h_{N+1-j},~~~~
\forall 1\le i<j\le N,~~\lambda_i,\lambda_j\in\sigma(\bfL).$$
Under this assumption we propose the following algorithm to construct a perfect reconstruction 
orthogonal filter bank from given parameters $\{y_i\}^s_{i=1}$.
\begin{alg}\label{alg:findyi}
Let $0=\lambda_1<\lambda_2\le...\le\lambda_N$ be all the eigenvalues of the graph Laplacian $\bfL$
and $r:=[\frac{N}{2}],~s:=[\frac{N+1}{2}]$.
\begin{enumerate}
\item 
Choose $2=y_1\ge...\ge y_s\ge 1$ and $y_s=1$ for odd $N$ satisfying
\begin{equation}\label{eq:y-ij-equal}
\lambda_i=\lambda_j~~\mbox{or}~~\lambda_{N+1-i}=\lambda_{N+1-j}
\implies y_i=y_j,~~~~i, j=1,..., s.
\end{equation}
\item
Set $y_{N+1-i}:=2-y_i,~i=1,...,s$. 
\item
For $i=1,...,N$, set $\bfg_0(i)=\bfh_0(i)=\sqrt{y_i}$ and $\bfg_1(i)=\bfh_1(i):=\bfg_0(N+1-i)$.
\item
Output the orthogonal filter bank: $\{\bfh_0, \bfg_0, \bfh_1, \bfg_1\}$.
\end{enumerate}
\end{alg}

\subsubsection{Mallat's Decomposition Algorithm}
As shown in Figure \ref{fig:2-channel-banks}, the input signal ${\bfx}^{(0)}$ is filtered to 
produce $\bfF_{h_0}{\bfx}^{(0)}$ and $\bfF_{h_1}{\bfx}^{(0)}$, which are further downsampled by
$\bfA_L$ and $\bfA_H$ to produce the following two shorter signals:
\begin{equation}\label{eq:decomposition}
\bfx^{(1)}:=\bfA_L\bfF_{h_0}{\bfx}^{(0)},~~~~
\bfz^{(1)}:=\bfA_H\bfF_{h_1}{\bfx}^{(0)},
\end{equation}
where $\bfA_L, \bfA_H$ are defined by \eqref{eq:ALAH}, namely,
$$\bfA_L=\frac{1}{\sqrt{2}}\bfU_1\bfP_0^{\top}{\bfU}^\top,~~~~~
\bfA_H=\frac{1}{\sqrt{2}}{\bfP}_1^{\top}{\bfU}^{\top}.$$
where ${\bfP}_0$ and ${\bfP}_1$ are designed according to \eqref{eq:P0-P1-def-even} and \eqref{eq:P0-P1-def-odd}.
\par 
According to Equation \eqref{eq:perfect-construction} for perfect reconstruction, we have
\begin{equation}\label{eq:reconstruction}
{\bfx}^{(0)}=\bfF_{g_0}\bfB_L\bfx^{(1)}
	+\bfF_{g_1}\bfB_H\mathbf{z}^{(1)},
\end{equation}
where $\bfB_L:=\bfA_L^\top$ and $\bfB_H:=\bfA_H^\top$.
\par
Given a signal $\bfx^{(0)}$ defined on $\mathcal{G}$, denote by $\bfx^{(1)}$ the output signal of the lowpass 
channel of the two-channel filter bank on $\mathcal{G}$. We equip $\bfx^{(1)}$ with 
a reduced graph $\mathcal{G}_1$ and design a new two-channel filter bank on $\mathcal{G}_1$ so that
$\bfx^{(1)}$ can be further decomposed.
The decomposition process can be implemented for several layers. 
The sequences of such decompositions and reconstructions, as described in \eqref{eq:decomposition} and \eqref{eq:reconstruction}, is illustrated in the flowchart in Table \ref{tab:Mallat-algorithm} and is called Mallat's algorithm.
\begin{table}[hbtp]
\caption{Mallat's algorithms for Decomposition and Reconstruction}
\begin{center}
\begin{tabular}{|c|c|}
\hline
Decomposition &
$\begin{array}{ccccccccc}
\bfx^{(0)} &\longrightarrow  & \bfx^{(1)}&\longrightarrow & \bfx^{(2)} &\longrightarrow &\cdots \\
       &\searrow & & \searrow& &\searrow \\[-3mm]
       & & \mathbf{z}^{(1)} &  &\mathbf{z}^{(2)} & &\cdots 
\end{array}$\\
\hline
Reconstruction &
$\begin{array}{ccccccccc}
\cdots &\longrightarrow & \bfx^{(2)}&\longrightarrow &\bfx^{(1)}&\longrightarrow &\bfx^{(0)}\\
&\nearrow&&\nearrow & &\nearrow \\[-3mm] 
\cdots & &\mathbf{z}^{(2)} && \mathbf{z}^{(1)} 
\end{array} $   \\
\hline    
\end{tabular}
\end{center}
\label{tab:Mallat-algorithm}
\end{table}

\section{Locality and Approximation Error}
\setcounter{equation}{0}
\label{sec:4}

\subsection{Locality of the Filters}
\subsubsection{Locality of the Filters: Theory}
\label{sec:locality}
Different requirements lead to different design of the filter banks. In some applications, one may want the filters to be well localized in the graph domain. 
In the classical signal processing, the key advantage of the wavelet transform compared 
to the Fourier Transform is the ability of extracting both local spectral and temporal information, which makes it very applicable for processing of non-stationary signals.  
In the classical two-channel filter bank, wavelet transforms serve as the analysis and synthesis filters, which corresponds to the analysis and synthesis filters
$\bfF_{h_0}, \bfF_{h_1}$ and $\bfF_{g_0}, \bfF_{g_1}$ in the graph settings as described in Section \ref{sec:TCFB}. Naturally, we concern about the locality of graph filters.
\par
Let us learn from the idea in \cite{narang2013compact} to characterize the locality of a filter $\bfF_h$. By
$$(\bfL\bfx)(i)=d_ix_i-\sum_{v_j\sim v_i}w_{ij}x_j,~~~~i=1,...,N,$$
where $v_j\sim v_i$ represents the edge connection between $v_i$ and $v_j$, 
we know that $(\bfL\bfx)(i)$ depends only on the 
values of the function $\bfx$ on the one-hop neighborhood of $v_i$: $\mathcal{N}(v_i):=\{
v_j|v_j\sim v_i\}$. The larger the weight $w_{ij}$, the greater the value $\bfx(v_j)$ contributes to $(\bfL\bfx)(i)$. Similarly,
$(\sum_{l=1}^k\bfL^l\bfx)(i)$ only depends on the $k$-hop neighborhood of $v_i$, where the $k$-hop neighborhood of the vertex $v_i$ refers to the set of vertices that can be connected to $v_i$ by at most $k$ edges. 
Therefore, if the filter $\bfF_h$ can be written as an
$m$-order polynomial in $\bfL$, then $(\bfF_h\bfx)(i)$ only depends on the signal 
values in the $m$-hop neighborhood of $v_i$. The degree $m$ of the  
polynomial can be regarded as an index of the locality of the filter $\bfF_h$. 
\par
We point out that the locality of graph filter $\bfF_h=p_m(\bfL)$ is consistent with the locality of 
traditional wavelet analysis. In the traditional case, if a filter
function of an orthonormal wavelet basis is a polynomial in the Fourier basis 
function $e^{-i\omega}$:
$$m_0(\omega)=\sum^m_{k=0}\xi_kz^k,~~~~z:=e^{-i\omega},$$ 
then the support of the scaling funtion $\phi$ and the wavelet function $\psi$ are respectively $\supp\phi\subset[0, m]$ and $\supp\psi\subset [1-m,m]$, and the Mallat's decomposition based on this wavelet basis is \cite{daubechies1992ten}:
$$y_k=\sqrt{2}\sum^m_{l=0}\xi_lx_{2k+l},~~~~
z_k=\sqrt{2}\sum^m_{l=0}(-1)^l\xi_lx_{2k+1-l}.$$
It can be seen that $y_k$ and $z_k$ depend only on the values on the vertices in the $(m+1)$-hop 
neighborhood of  $x_{2k}$.
\par
The perfect reconstruction filters $\bfF_{h_0}$ and $\bfF_{h_1}$ designed according to the proposed method need to meet the conditions \eqref{eq:th:PR-cond2-a}. It is usually difficult to find polynomials $h_0,h_1,g_0,g_1$ in $\lambda$ such that 
\eqref{eq:th:PR-cond2-a} holds for
\[
\bfh_i=(h_i(\lambda_1),...,h_i(\lambda_N))^\top,~~\bfg_i=(g_i(\lambda_1),...,g_i(\lambda_N))^\top,~~i=0,1,
\]
where $\lambda_i\in\sigma(\bfL)$. 
However, if the vectors $\bfh_0$ and $\bfh_1$ that satisfy \eqref{eq:th:PR-cond2-a} can be approximated by polynomials in $\lambda\in\br$ on $\sigma(\bfL)$, then $\bfF_{h_0}$ and $\bfF_{h_1}$ are said to be approximately localized. This is explained by the following theorem.
\begin{thm}\label{th:approx-by-poly}
Denote the eigenvalues of the graph Laplacian matrix $\bfL$ by $0=\lambda_1\le...\le\lambda_N$. Then for any $ \bfh\in\{\bfh\in\br^N~|~h_i=h_j~if~\lambda_i=\lambda_j\}$, there exists an $m$-order polynomial $p_m$, such that
$$\|\bfF_h-p_m(\bfL)\|_2\le\frac{6\lambda_N}{m}M_h,$$
where $M_h$ is the Lipschitz constant of the filter vector $\bfh$ defined by
$$M_h:=\max_{1\le i\le N-1\atop\lambda_{i+1}\neq\lambda_i}
	\Big|\frac{h_{i+1}-h_i}{\lambda_{i+1}-\lambda_i}\Big|.$$
\end{thm}
\prf
Suppose the eigen decomposition of $\bfL$ is $\bfL=\bfU\Lambda \bfU^{\top}$, where $\Lambda:=\diag(\lambda_1,...,\lambda_N)$, then
\begin{align*}
	\|\bfF_h-p_m(\bfL)\|_2
	&=\|\bfU[\diag(\bfh)-p_m(\Lambda)]\bfU^{\top}\|_2
	=\|\diag(\bfh)-p_m(\Lambda)\|_2\\
	&=\max_{1\le i\le N}|h_i-p_m(\lambda_i)|.
\end{align*}
\par
Let $f$ be a piecewise function connecting all the points $\{(\lambda_i,h_i)\}^N_{i=1}$ in turn, then
$f\in\mathrm{Lip}_{M_h}1$. Namely, $f$ is a function on $[0,\lambda_N]$ that satisfies the following Lipschitz condition:
$$|f(x)-f(y)|\le M_h|x-y|,~~~~\forall x,y\in [0,\lambda_N].$$
According to \cite[Corollary 1, \S 6.2]{natanson1961constructive}, there is an $m$-order polynomial $p_m$ 
such that
$$\max_{\lambda\in [0,\lambda_N]}|f(\lambda)-p_m(\lambda)|\le\frac{6\lambda_N}{m}M_h.$$
Since $h_i=f(\lambda_i),~i=1,...,N$, we have 
$\|\bfF_h-p_m(\bfL)\|_2\le\frac{6\lambda_N}{m}M_h$.
\bbox
\par
In order to construct an orthogonal filter bank $\{\bfh_0, \bfg_0, \bfh_1, \bfg_1\}$, 
according to Algorithm \ref{alg:findyi}, we only 
need to choose $2=y_1\ge...\ge y_s\ge 1$ that satisfies \eqref{eq:y-ij-equal}. The Lipschitz constants
of these filters are all equal to
$$M:=\max_{1\le i\le N-1\atop\lambda_{i+1}\neq\lambda_i}
	\Big|\frac{\sqrt{y_i}-\sqrt{y_{i+1}}}{\Delta\lambda_i}\Big|,~~~~
\mbox{with}~~\Delta\lambda_i:=\lambda_{i+1}-\lambda_i.$$
To make $\bfh_0=\bfg_0$ and $\bfh_1=\bfg_1$ have as best locality as possible, by Theorem 
\ref{th:approx-by-poly}, we want the constant $M$ to be as small as possible.
When $N$ is odd, we have $y_{s}=2-y_s$, which implies that $y_s=1$. For simplicity, we also set 
$y_s=1$ when $N$ is even, which yields $y_{r+1}=2-y_s=1$. Therefore, it always holds that  
$y_s=y_{r+1}=1$ no matter $N$ is even or odd.
\par
For $i=1,...,s-1$, let $\{\alpha_i\}$ and $\{\beta_i\}$ be nonnegative numbers satisfying
\begin{equation}\label{eq:sqrt-y-alpha-beta}
	\begin{cases}
		\sqrt{y_i}-\sqrt{y_{i+1}}=\alpha_i\Delta\lambda_i,\\
		\sqrt{2-y_{i+1}}-\sqrt{2-y_i}=\beta_i\Delta\lambda_{N-i},\end{cases}
		~~i=1,...,s-1.
\end{equation}
It is easy to see that $\{y_i\}^s_{i=1}$ is determined uniquely by $\{\alpha_i\}^{s-1}_{i=1}$ or 
$\{\beta_i\}^{s-1}_{i=1}$ given $y_1=2$.
Summing both sides of \eqref{eq:sqrt-y-alpha-beta} for $i$ from $1$ to $s-1$, we get the following 
constraints of $\bm{\alpha}:=[\alpha_1,\cdots,\alpha_{s-1}]$ and $\bm{\beta}:=[\beta_1,...,\beta_{s-1}]$:
\begin{equation}\label{eq:constrain-alpha-beta}
	\begin{cases}
		\sum^{s-1}_{i=1}\alpha_i\Delta\lambda_i=\sqrt{2}-1,\\
		\bm{\alpha}\succeq 0,\end{cases}~~~~
	\begin{cases}
		\sum^{s-1}_{i=1}\beta_i\Delta\lambda_{N-i}=1,\\
		\bm{\beta}\succeq 0.\end{cases}
\end{equation}
Based on the constraints we propose two strategies for constructing  $\bm{\alpha}$ or $\bm{\beta}$
as follows:
\begin{itemize}
	\item Find $\bm{\alpha}\in\br_+^{s-1}$ which minimizes $\|\bm{\alpha}\|_\infty$ 
	under the first constraint of \eqref{eq:constrain-alpha-beta}.
	\item
	Find $\bm{\beta}\in\br_+^{s-1}$ which minimizes $\|\bm{\beta}\|_\infty$ under the second constraint of \eqref{eq:constrain-alpha-beta}.
\end{itemize}
\par
For the sake of simplicity, we only describe the first strategy here. The another one is similar. The minimum solution of $\|\bm{\alpha}\|_\infty$ under the first 
constraint of \eqref{eq:constrain-alpha-beta} can be obtained by the following lemma.
\begin{lem}\label{lem:optimization-infty}
Given nonzero $\mathbf{a}=(a_1,...,a_n)^\top\in\br_+^n$ and $b\in\br$, if 
the optimization problem
\begin{equation}\label{eq:lem:optimization-infty}
\min\|\bfx\|_\infty,~~~~\st~~\mathbf{a}^{\top}\bfx=b,~~\bfx\succeq \mathbf{0}
\end{equation}
has nonempty feasible set
$\mathcal{D}:=\{\bfx\in\br^n|\mathbf{a}^{\top}\bfx=b,~\bfx\succeq \mathbf{0}\}$,
then the optimal solution $x^*$ exists and satisfies 
\begin{equation}\label{eq:lem:optimization-infty-1}
x^*_i=\|\bfx^*\|_\infty,~~~~\forall i\in I:=\{1\le i\le n|a_i\neq 0\}.
\end{equation}
\end{lem}
\prf
Since $\mathcal{D}$ is not an empty set, it is easy to show that the infimum of 
$f(\bfx):=\|\bfx\|_\infty$ on $\mathcal{D}$ is reachable at some point 
$\bfx^*\in\mathcal{D}$, which is a solution of the optimization problem
\eqref{eq:lem:optimization-infty}.
\par
Next, let us prove \eqref{eq:lem:optimization-infty-1} for any solution
$\bfx^*$ of \eqref{eq:lem:optimization-infty}.
\par
We assume that $I$ is not empty without losing generality. If 
\eqref{eq:lem:optimization-infty-1} is not true, there must exist a $k\in I$ such that
$x_k^*<\|\bfx^*\|_\infty$. For any $\epsilon>0$, using $a_k\neq 0$ we have that
$$b=\sum_{i\in I}a_ix_i^*
=a_k\Big(x_k^*+a_k^{-1}\epsilon\sum_{i\in I\setminus\{k\}}a_ix_i^*\Big)
+\sum_{i\in I\setminus\{k\}}a_i(x_i^*-\epsilon x_i^*).$$
Define $\tilde{\bfx}:=(\tilde{x}_1,..., \tilde{x}_n)^{\top}$ as follows:
$$\tilde{x}_i:=\begin{cases}
x_k^*+a_k^{-1}\epsilon\sum_{i\in I\setminus\{k\}}a_ix_i^* & i=k,\\
(1-\epsilon)x^*_i & i\in I\setminus\{k\},\\
0 & i\notin I.\end{cases}
$$
It is easy to see that, for $\epsilon>0$ sufficiently small, there holds
$\tilde{\bfx}\succeq 0$, $\mathbf{a}^{\top}\tilde{\bfx}=b$, and
$|\tilde{x}_k|<\|\bfx^*\|_\infty$. Thus, for any $i\in I\setminus\{k\}$, we have
$$|\tilde{x}_i|=(1-\epsilon)|x^*_i|\begin{cases}
=0 <\|\bfx^*\|_\infty& x^*_i=0,\\
<|x^*_i|\le\|\bfx^*\|_\infty&x^*_i\neq 0,
\end{cases}$$
and consequently $\|\tilde{\bfx}\|_\infty<\|\bfx^*\|_\infty$, which contradicts 
the assumption that $\bfx^*$ is an optimal solution.
\bbox
\par
By Lemma \ref{lem:optimization-infty} we can find an $\bm{\alpha}\in\br^{s-1}_+$ according to the 
first strategy, which is
$$\alpha_1=\cdots=\alpha_{s-1}=\frac{\sqrt{2}-1}{\sum^{s-1}_{i=1}\Delta\lambda_i}
=\frac{\sqrt{2}-1}{\lambda_s}.$$
Inserting it into \eqref{eq:sqrt-y-alpha-beta}, we obtain that
\begin{align}\label{eq:alpha_formula}
	y_i=\Big[\sqrt{2}-(\sqrt{2}-1)\frac{\lambda_i}{\lambda_s}\Big]^2,~~~~i=1,... s.
\end{align}
\par
Similarly, according to the second strategy we can find $\bm{\beta}\in\br^{s-1}_+$ as
$$\beta_1=\cdots=\beta_{s-1}=\frac{1}{\lambda_N-\lambda_{r+1}}.$$
Consequently, $\{y_i\}^s_{i=1}$ are as follows:
\begin{equation}
y_i=2-\Big(\frac{\lambda_N-\lambda_{N+1-i}}{\lambda_N-\lambda_{r+1}}\Big)^2,
 ~~~~ i=1,... s.
\end{equation}
\par
The Lipschitz constants of the filter vector desigen by the two above strategies 
may be different. In practical applications, we can choose the one with smaller 
Lipschitz constant.
\par
To verify the above theory on the locality of the filters, let us construct the following
two typical examples of perfect reconstruction orthogonal  two-channel filter banks 
by Algorithm \ref{alg:findyi}:
\begin{itemize}
\item \textbf{localFB}
\begin{align*}
y_i=\Big[\sqrt{2}-(\sqrt{2}-1)\frac{\lambda_i}{\lambda_s}\Big]^2,~~i=1,... s;~~~~
y_i=2-y_{N-i+1},~~i=s+1,...,N.
\end{align*}
\item \textbf{idealFB}
$$y_1=...=y_{s-1}=2,~~~y_{s+1}=...=y_N=0,~~~y_s=\begin{cases}
1 & \ifs N~\mbox{is odd},\\
2 & \ifs N~\mbox{is even}.\end{cases}
$$
\end{itemize}
After getting $\{y_i\}^N_{i=1}$, we calculate filters as follows:
$$\bfg_0(i)=\bfh_0(i)=\sqrt{y_i},~~~\bfg_1(i)=\bfh_1(i):=\bfg_0(N+1-i),~~~~
i=1,...,N.$$
\par
\begin{figure}[hbtp]
\centerline{
\includegraphics[width=\textwidth,height=5cm]{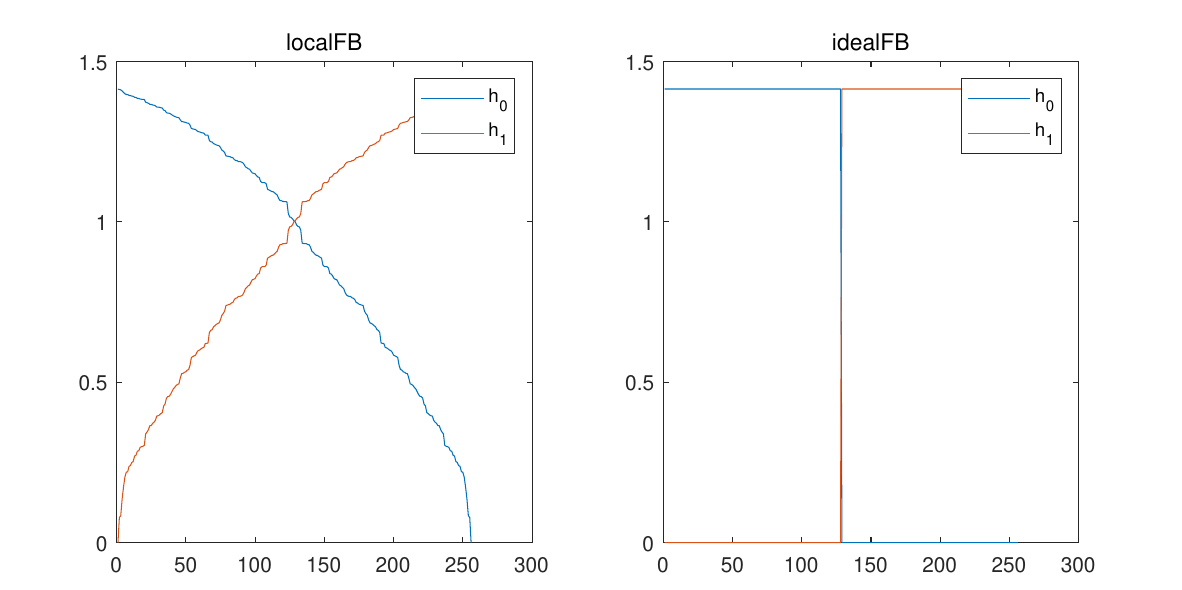}}
\caption{\small  Two types of orthogonal filter banks constructed using the proposed method. Left: localFB, right: idealFB.}
\label{fig:filter-eg}
\end{figure}

The localFB depends on the eigenvalues of the graph Laplacian, 
while the idealFB depends only on the number of vertices of the graph.
Figure \ref{fig:filter-eg} shows the filters $\bfh_0$ and $\bfh_1$ for 
localFB and idealFB for the ring graph with $256$ vertices.
\par
\begin{figure}[hbtp]
\centerline{
\includegraphics[scale=0.3]{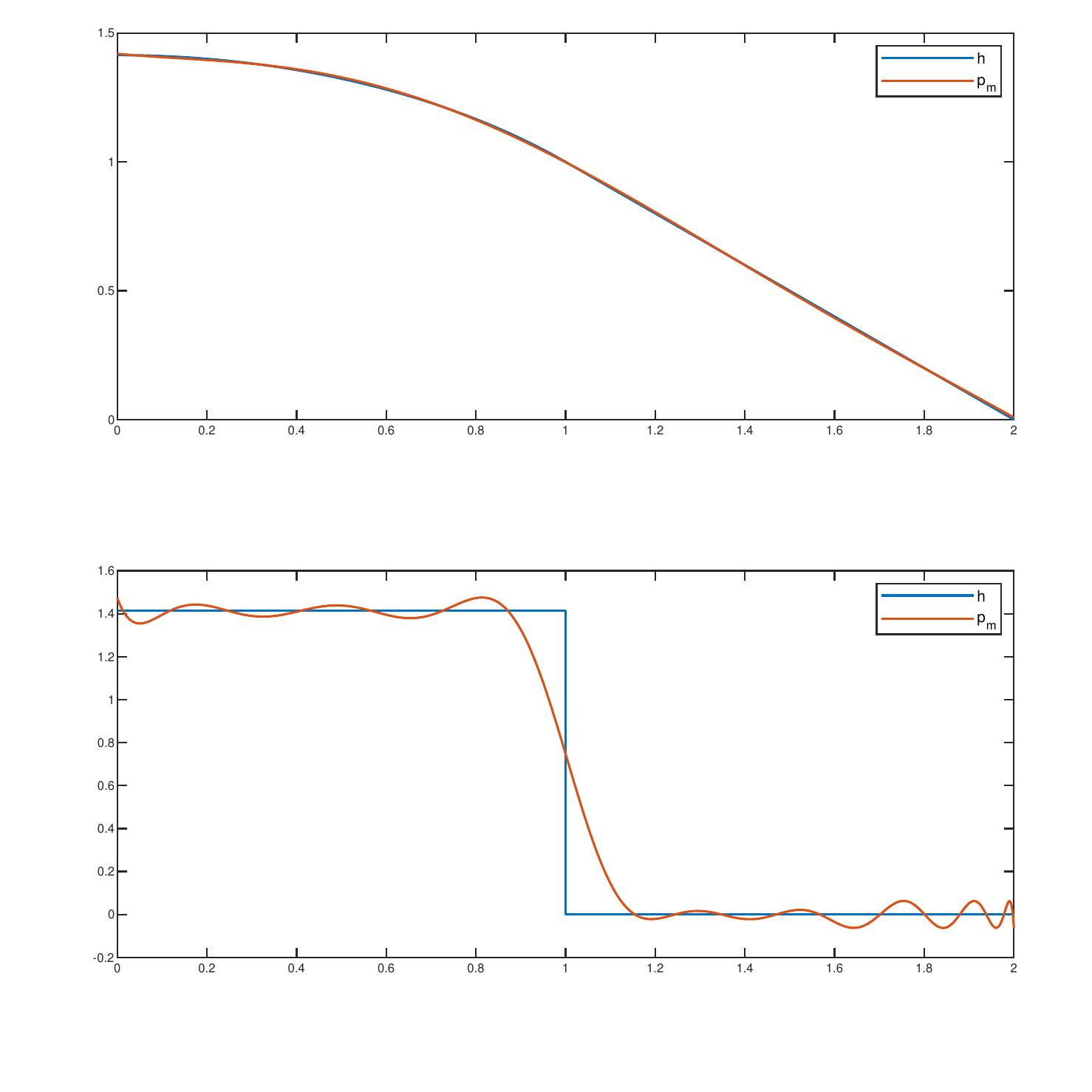}
\includegraphics[scale=0.3]{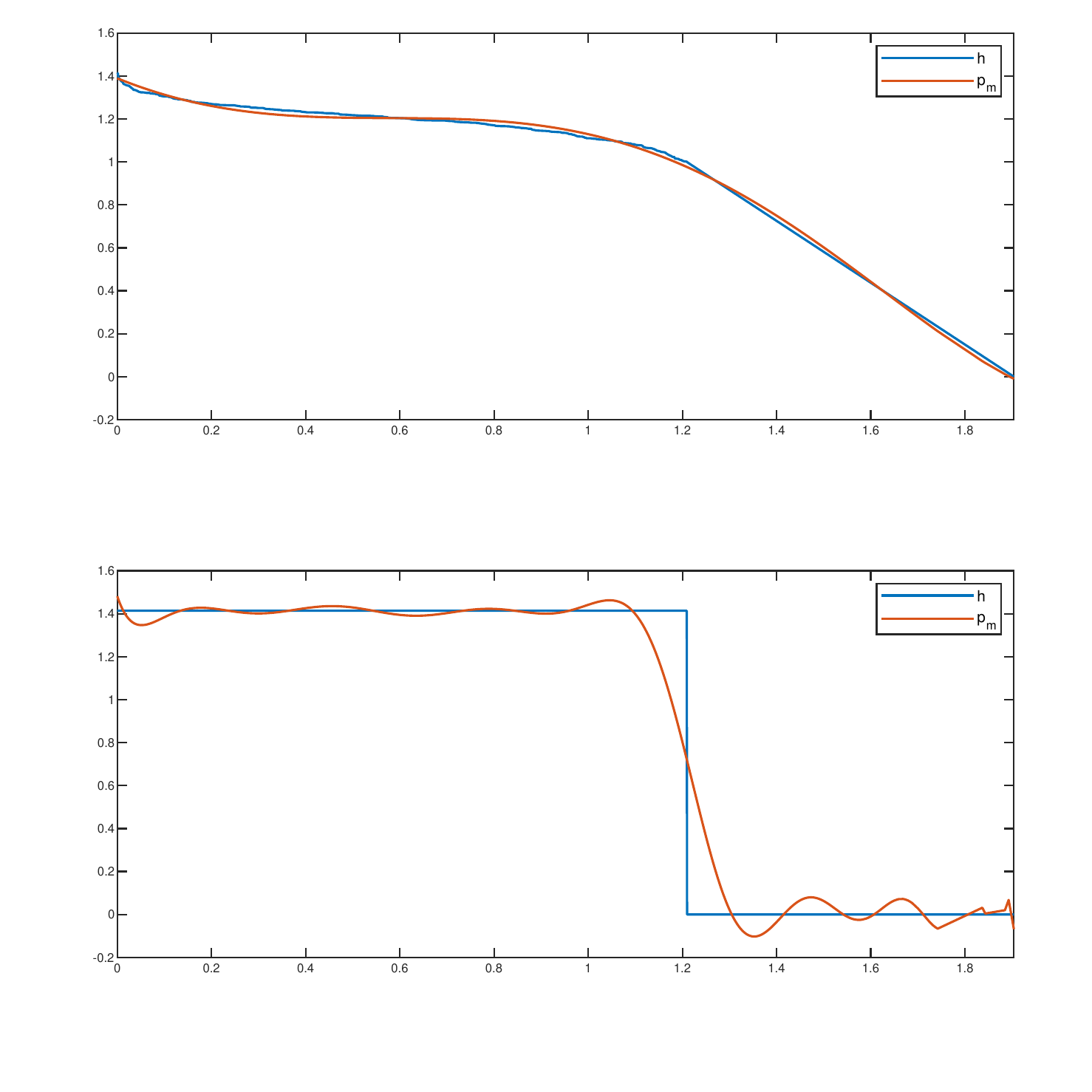}}
\caption{\small The top row shows the filter functions $h$'s designed by the proposed localFB
	and their $5$th-order uniform approximation polynomials $p_m$ on the ring graph (left) and on the 
	sensor graph (right) with $1000$ vertices. The bottom row correspondingly shows the ideal half-band filter functions $h$'s and their  $30$th-order uniform approximation polynomials.}
\label{fig:filter-lowM}
\end{figure}

Figure  \ref{fig:filter-lowM}\footnote{	Since  calculating the best approximation in the uniform norm is computationally difficult, high-degree approximation polynomial for the ideal half-band filter may not be very accurate.}
shows the filter functions $h(\lambda_i):=\bfh(i),~i=1,...,N$ (blue color) 
and their uniform approximation polynomial $p_m(\lambda)$ (red color) on the ring 
graph (left) and the sensor graph (right) with $1000$ vertices. The two filter functions 
on the top row are produced by the proposed localFB (the one with smaller
Lipschitz constant). Their  Lipschitz constants 
$M$'s are respectively $1$ (ring graph) and $1.4396$ (sensor graph).
By the Remez algorithm \cite{remes1934calcul}, the corresponding $5$th-order best uniform approximation polynomials
are shown as the red curve. 
For comparison, the ideal half-band filters of idealFB and their corresponding $30$th-order best uniform approximation
polynomials are shown in the bottom row of the figure. The Lipschitz constants 
$M$'s, which are respectively $224.478$ (ring graph) and $533.7062$ (sensor graph), are much larger than those of localFB, so the ideal half-band filters cannot be well approximated uniformly by low-order polynomials.

\subsubsection{Locality of the Filters: Experiments}
\begin{figure}[hbtp]
\centerline{
\includegraphics[width=0.3\textwidth,height=0.15\textheight]{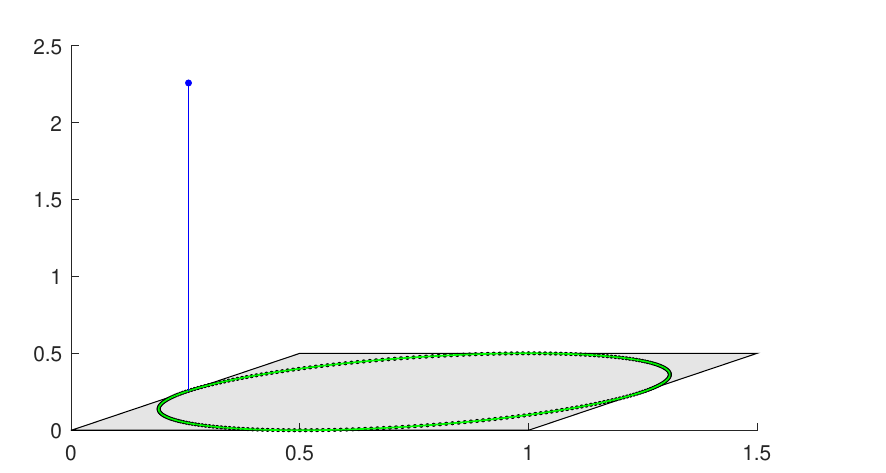}
\includegraphics[width=0.3\textwidth,height=0.15\textheight]{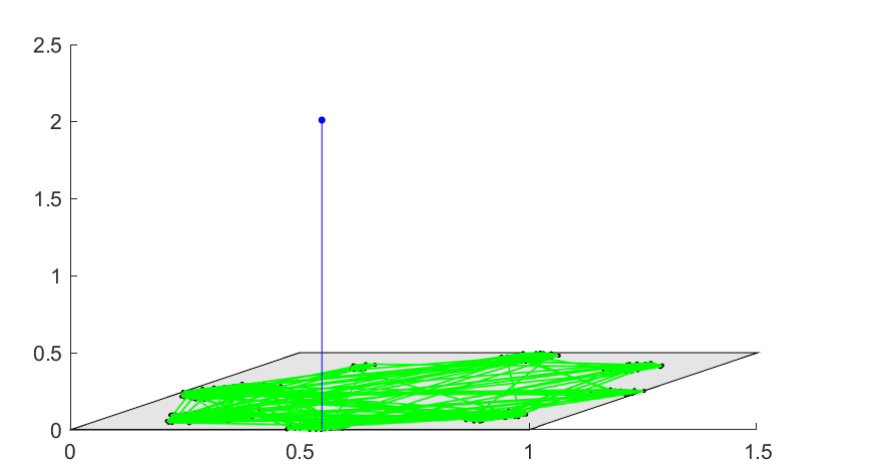}
\includegraphics[width=0.3\textwidth,height=0.15\textheight]{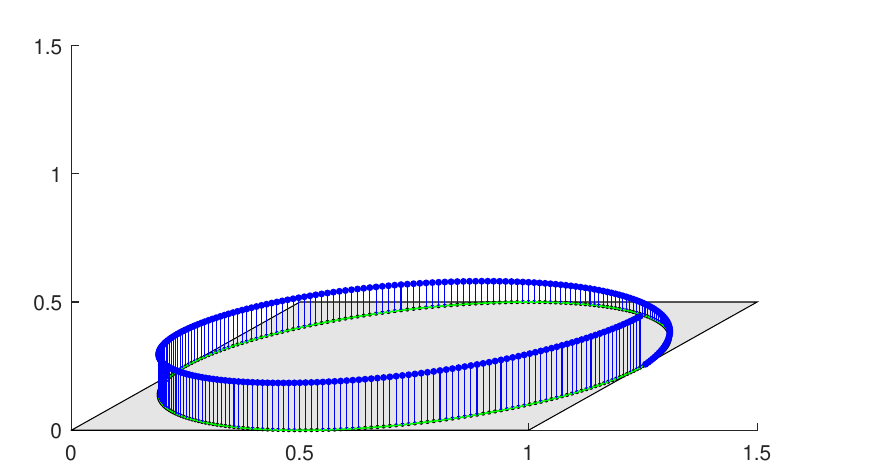}
}
\caption{\footnotesize 
From left to right are the impulse signals on the ring graph, the community 
network of $256$ vertices, and the discontinuous signal defined by
\eqref{eq:step-signal} on the ring graph.
}
\label{fig:impulse signal}
\end{figure}

In this section, we conduct experiments to verify the locality of this 
two types of filters: localFB and idealFB designed in \ref{sec:locality}. We consider
the impulse signals on the ring graph and the community network, both have
$256$ vertices, and a discontinuous signal 
\begin{equation}\label{eq:step-signal}
\bfx(k)=0.2\sin\Big(\frac{k-1}{2(N-1)}\pi\Big),~~~~k=1,...,N,
\end{equation}
on the ring graph with $N=256$ vertices, which contains a step between the first vertex 
and the last one. The signals described above are shown in Figure \ref{fig:impulse signal}.
\par
\begin{figure}[hbtp]
\centerline{
\includegraphics[scale=0.4]{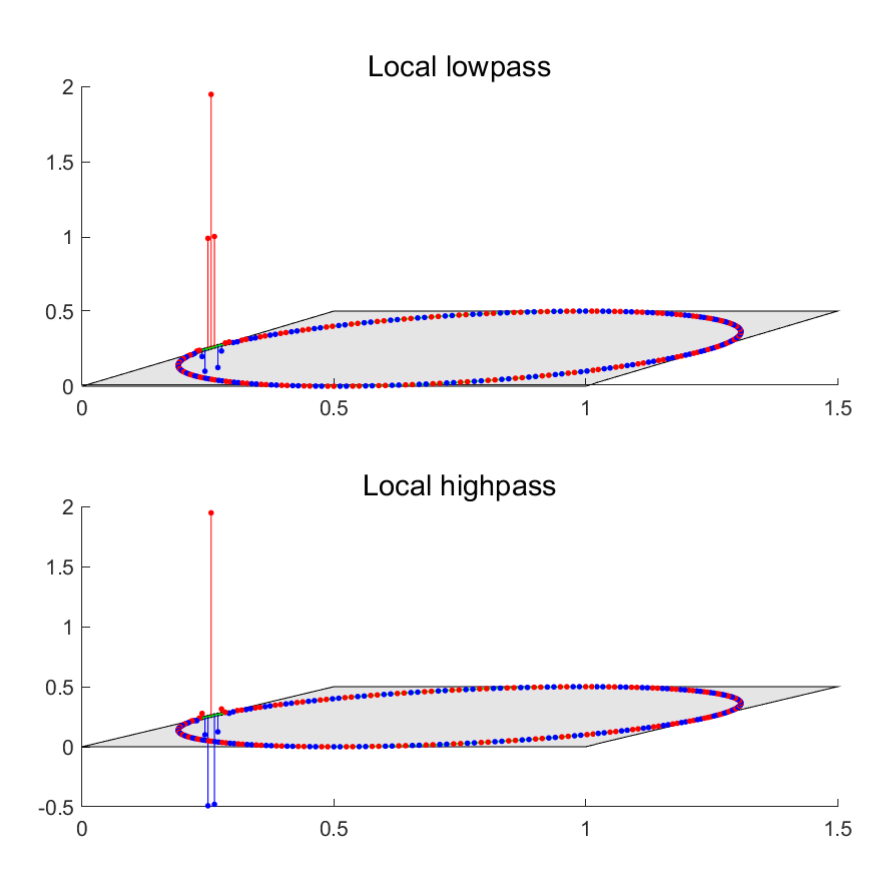}
\includegraphics[scale=0.4]{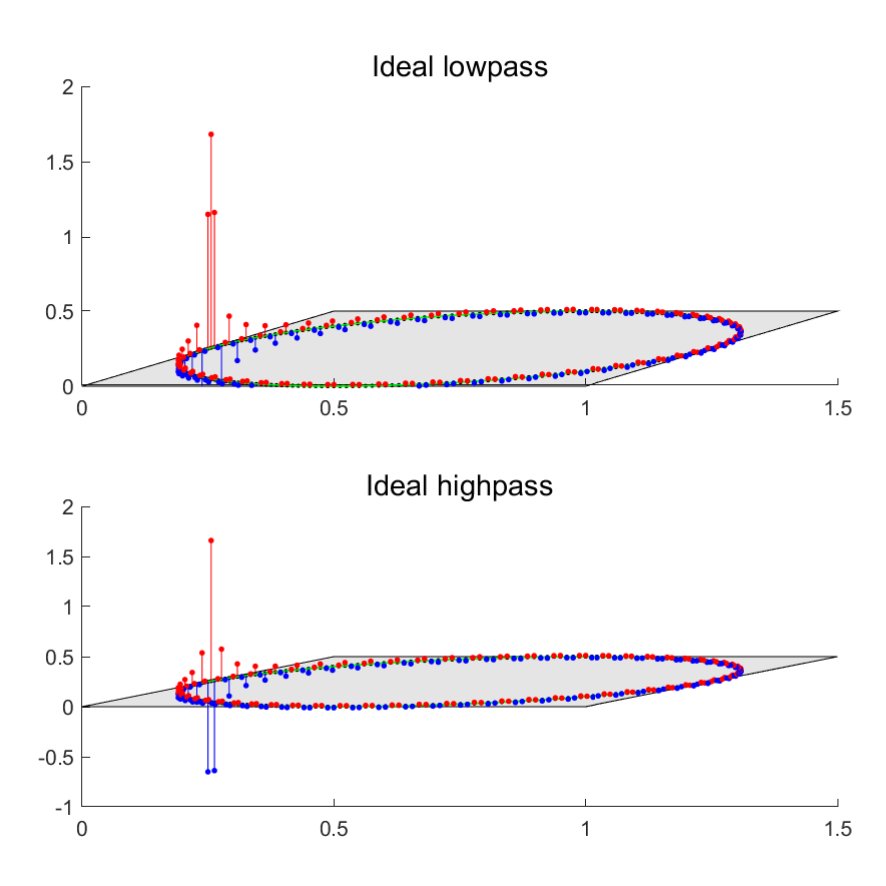}}
\caption{\footnotesize 
Filtered signals $\bfF_{h_0}\bfx$ and $\bfF_{h_1}\bfx$ for the impulse 
signal $\bfx$ on the ring graph by localFB (left) and idealFB (right). }
\label{fig:ring_id_lo}
\end{figure}
\begin{figure}[hbtp]
\centerline{
\includegraphics[scale=0.4]{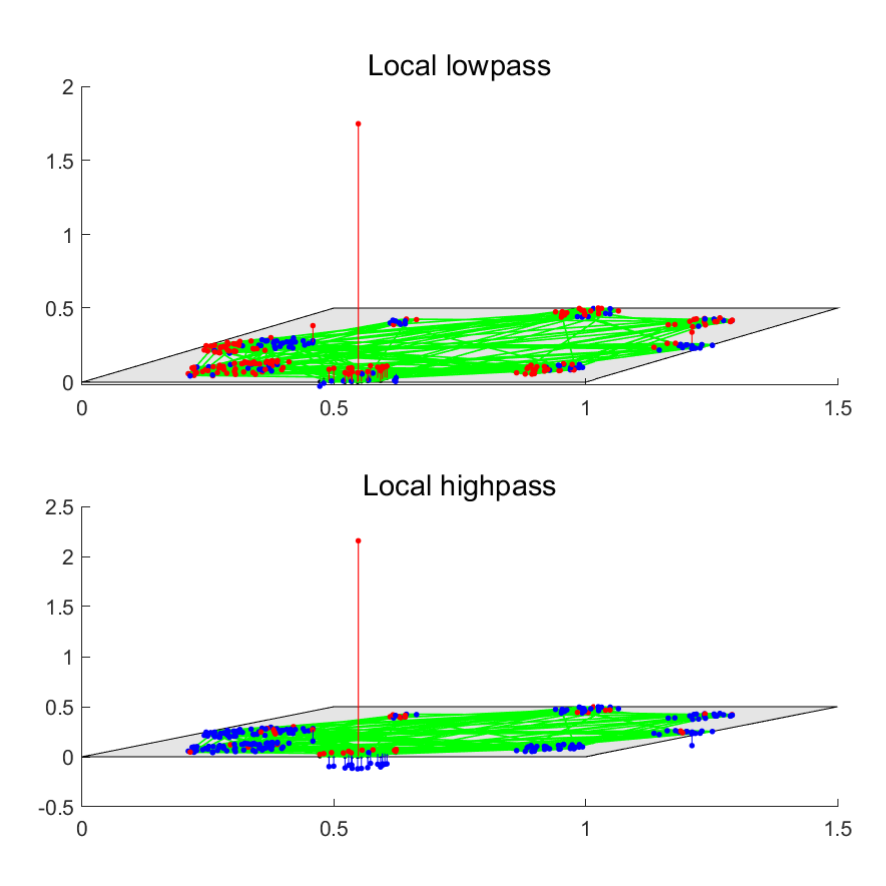}
\includegraphics[scale=0.4]{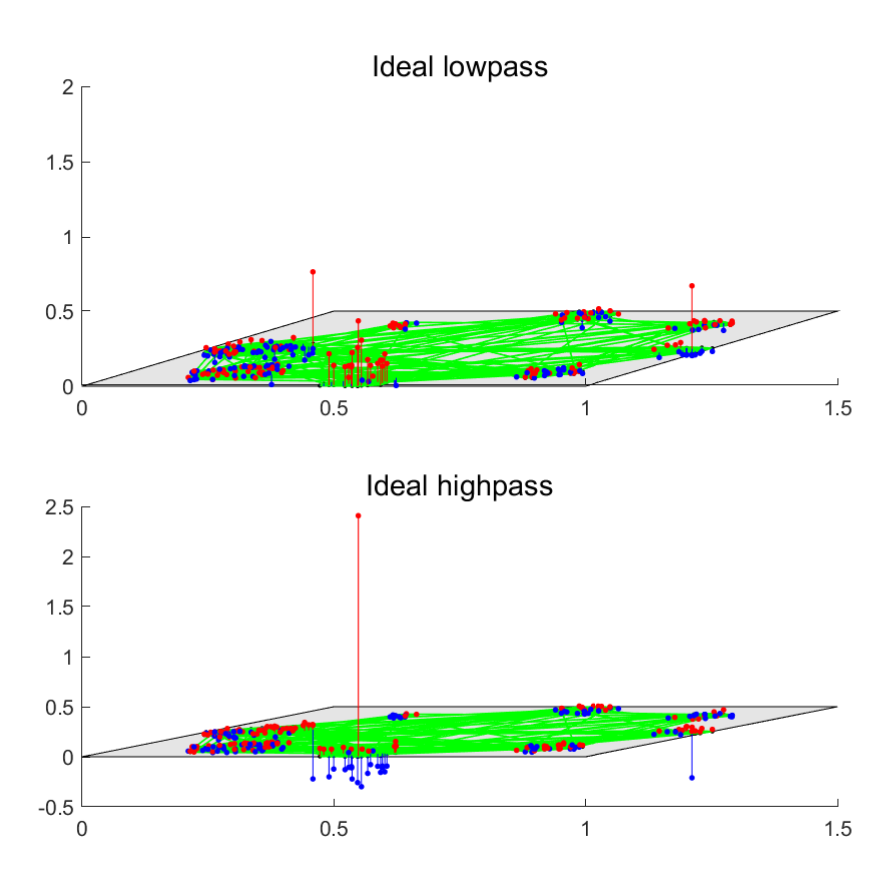}
}
\caption{\footnotesize 
Filtered signals $\bfF_{h_0}\bfx$ and $\bfF_{h_1}\bfx$ for the impulse 
signal $\bfx$ on the community graph by localFB (left) and idealFB (right). }\label{fig:commu_id_lo}
\end{figure}
\begin{figure}[hbtp]
\centerline{
\includegraphics[scale=0.4]{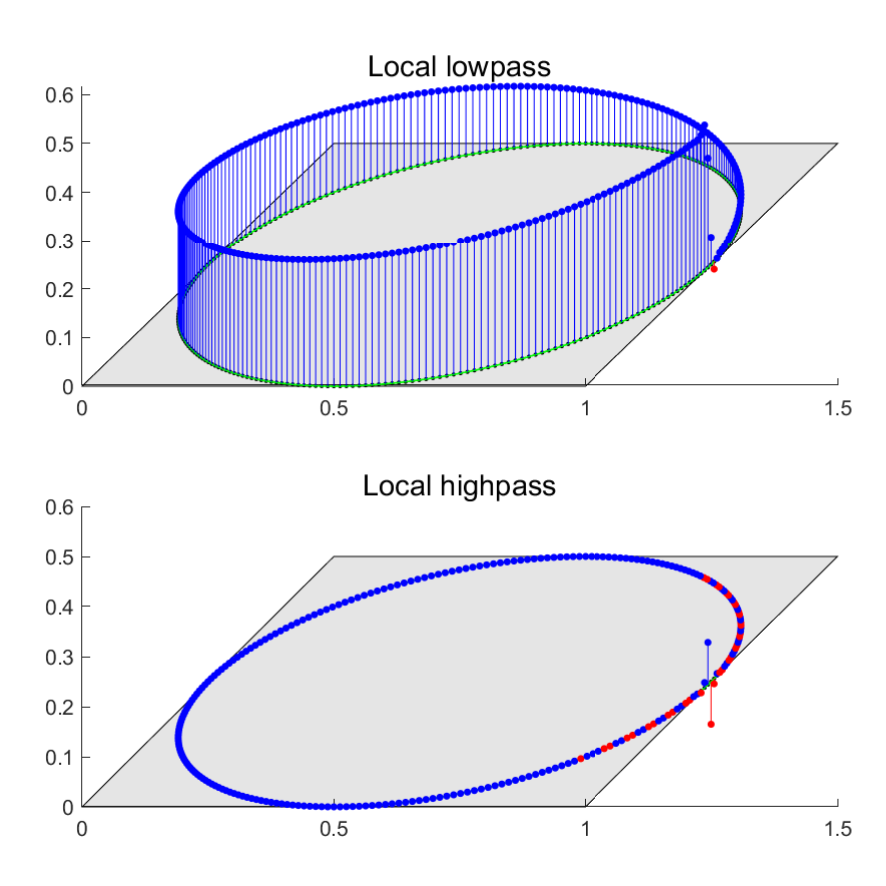}
\includegraphics[scale=0.4]{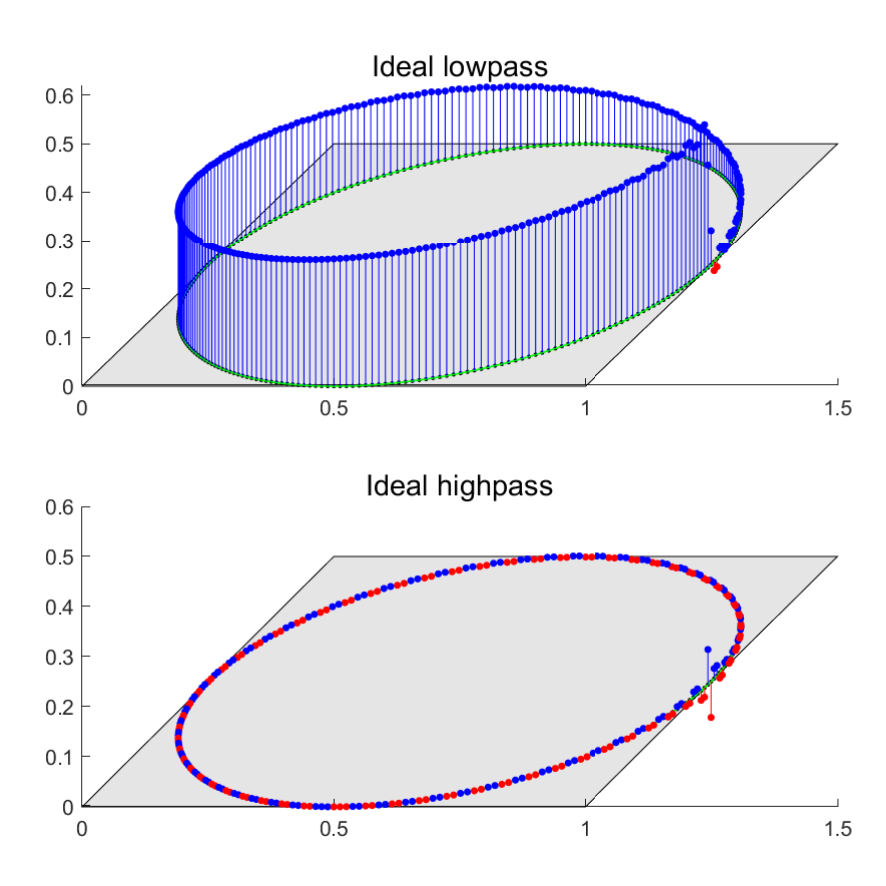}
}
\caption{\footnotesize 
Filtered signals $\bfF_{h_0}\bfx$ and $\bfF_{h_1}\bfx$ for the discontinuous signal 
$\bfx$ defined by \eqref{eq:step-signal} on the ring graph by localFB (left) and idealFB (right). }
\label{fig:ring_step_LH}
\end{figure}
\begin{figure}[hbtp]
\centerline{
\includegraphics[width=0.4\textwidth,height=0.3\textheight]{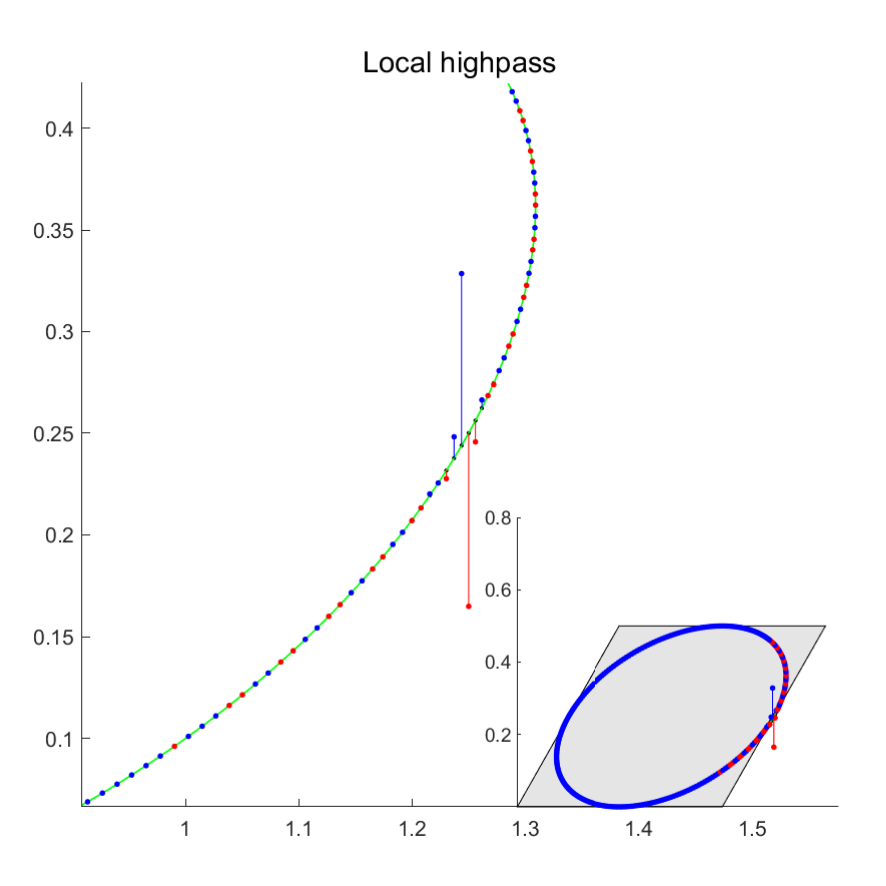}
\includegraphics[width=0.4\textwidth,height=0.3\textheight]{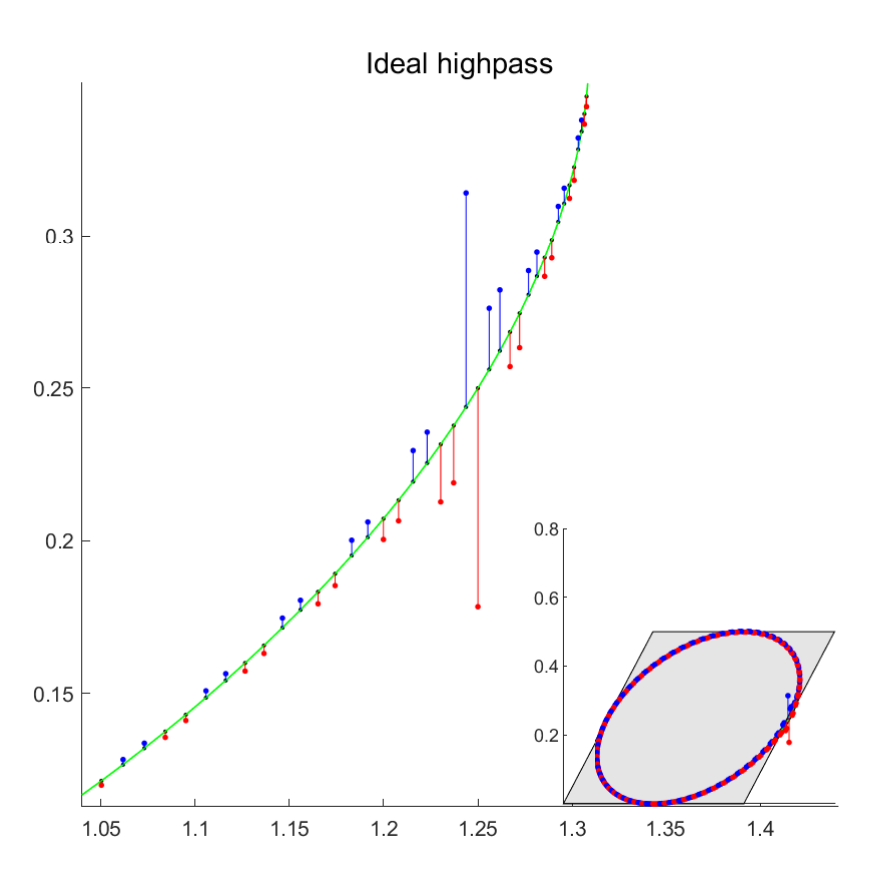}
}
\caption{\footnotesize 
Partial enlargement near the step point in the figures in the bottom row of Figure \ref{fig:ring_step_LH}. }
\label{fig:ring_step}
\end{figure}

We filter the above three signals with $\bfh_0$ (lowpass) and $\bfh_1$ (highpas) 
of localFB and idealFB and show the experimental results in 
Figures \ref{fig:ring_id_lo}, \ref{fig:commu_id_lo} and \ref{fig:ring_step_LH}.
All experiments in this section and the next section are done with Matlab, and the toolbox involved is mainly GSPBox for matlab \cite{perraudin2014gspbox}. It shows that, using idealFB, the filtered signals $\bfF_{h_0}\bfx$ and $\bfF_{h_1}\bfx$ have widespread oscillations around the discontinuities, that is,  samples in a wide range around the impulse/step are badly affected. In contrast, in the filtered signals of localFB, only samples in a narrow range around the impulse/step are affected. 
This phenomenon can be observed more clearly by enlarging the part near the 
step point in Figure \ref{fig:ring_step_LH}, as shown in Figure \ref{fig:ring_step}.
These experiments validate our theoretic conclusion: the filters of localFB is of much better 
locality in the vertex domain than those of idealFB. 

\subsection{Approximation Error}
\subsubsection{Approximation Error: Theory}
For a smooth signal $\bfx$, it is expected that the reconstructed signal using only the output of the lowpass channel can approximate $\bfx$, i.e.,  $\bfF_{g_0}\bfB_L\bfA_L\bfF_{h_0}\bfx\approx \bfx$. By \eqref{eq:BA-by-Q} and \eqref{eq:ch2-PR-tmp1}, 
we have 
$$\bfI_N-\bfF_{g_0}\bfB_L\bfA_L\bfF_{h_0}
=\bfI_N-\frac{1}{2}\bfF_{g_0}(\bfI+\bfQ)\bfF_{h_0}
=\frac{1}{2}\bfF_{g_1}(\bfI-\bfQ)\bfF_{h_1}.$$
Since $\bfF_{h_1}\bfQ=\bfQ\bfF_{h_1\circ\kappa}$ and $\bfQ$ is symmetric, 
where $\kappa(k)=N+1-k$, 
it follows that
$$\bfI_N-\bfF_{g_0}\bfB_L\bfA_L\bfF_{h_0}
=\frac{1}{2}\bfF_{g_1}\big(\bfF_{h_1}-\bfF_{h_1\circ\kappa}\bfQ\big).$$
Therefore, for any $\bfx\in\br^N$, there holds
\begin{align*}
\bfF_{g_1}\big(\bfF_{h_1}-\bfF_{h_1\circ\kappa}\bfQ\big)\bfx
&=\bfU\diag(\bfg_1)\big[\diag(\bfh_1)\hat{\bfx}-\diag(\bfh_1\circ\kappa)\Phi\hat{\bfx}\big]\\
&=\bfU\diag(\bfg_1)\big[\diag(\bfh_1)\hat{\bfx}-\diag(\bfh_1\circ\kappa)(\hat{\bfx}\circ\kappa)\big],
\end{align*}
where $\Phi=\bfU^{\top}\bfQ\bfU$. Hence,
\begin{align*}
\|\bfF_{g_1}\big(\bfF_{h_1}-\bfF_{h_1\circ\kappa}\bfQ\big)\bfx\|_2^2
&=\sum^N_{i=1}|\bfg_1(i)|^2|\bfh_1(i)\hat{\bfx}(i)-\bfh_1(N+1-i)\hat{\bfx}(N+1-i)|^2\\
&=\sum^N_{i=1}|\bfh_0(N+1-i)|^2|\bfg_0(N+1-i)\hat{\bfx}(i)-\bfg_0(i)\hat{\bfx}(N+1-i)|^2.
\end{align*}
\par
If $N$ is odd, we have $s=r+1$, $N+1-s=s$ and
$$|\bfg_1(s)|^2|\bfh_1(s)\hat{\bfx}(s)-\bfh_1(N+1-s)\hat{\bfx}(N+1-s)|^2=0.$$
If $N$ is even, there holds $s=r$ and the above term does not exist.
In both cases we have 
\begin{align*}
\|&\bfF_{g_1}\big(\bfF_{h_1}-\bfF_{h_1\circ\kappa}\bfQ\big)\bfx\|_2^2\\
&=\sum^r_{i=1}|\bfh_0(N+1-i)|^2|\bfg_0(N+1-i)\hat{\bfx}(i)-\bfg_0(i)\hat{\bfx}(N+1-i)|^2\\
&+\sum^r_{j=1}|\bfh_0(j)|^2|\bfg_0(j)\hat{\bfx}(N+1-j)-\bfg_0(N+1-j)\hat{\bfx}(j)|^2\\
&=\sum^r_{i=1}\big[|\bfh_0(i)|^2+|\bfh_0(N+1-i)|^2\big]
|\bfg_0(N+1-i)\hat{\bfx}(i)-\bfg_0(i)\hat{\bfx}(N+1-i)|^2\\
&=\sum^r_{i=1}|c_i(\bfh_0)|^2|\bfg_0(N+1-i)\hat{\bfx}(i)-\bfg_0(i)\hat{\bfx}(N+1-i)|^2,
\end{align*}
where
$$c_i(\bfh_0):=\sqrt{|\bfh_0(i)|^2+|\bfh_0(N+1-i)|^2}.$$
It is easy to see that $c_{N+1-i}(\bfh_0)=c_i(\bfh_0)$. 
If $\bfg_0(N)=0$, using the Minkowski's inequality we obtain that
\begin{align*}
\|\bfF_{g_1}\big(\bfF_{h_1}-\bfF_{h_1\circ\kappa}\bfQ\big)\bfx\|_2
&\le\Big(\sum^r_{i=1}|c_i(\bfh_0)\bfg_0(N+1-i)\hat{\bfx}(i)|^2\Big)^{1/2}
+\Big(\sum^r_{i=1}|c_i(\bfh_0)\bfg_0(i)\hat{\bfx}(N+1-i)|^2\Big)^{1/2}\\
&\le A_1\sigma_1(\bfx)^{1/2}+ A_2\sigma_2(\bfx)^{1/2},
\end{align*}
where
\begin{equation}\label{eq:-sigma-A}
\begin{cases}
\sigma_1(\bfx):=\sum^r_{i=1}\lambda_i|\hat{\bfx}(i)|^2,\\
\sigma_2(\bfx):=\sum^N_{i=s+1}\lambda_i|\hat{\bfx}(i)|^2,
\end{cases}~~~~
\begin{cases}
A_1:=\max_{2\le i\le r}\lambda_i^{-1/2}|c_i(\bfh_0)\bfg_0(N+1-i)|;\\
A_2:=\max_{s+1\le i\le N}\lambda_i^{-1/2}|c_i(\bfh_0)\bfg_0(N+1-i)|.
\end{cases}
\end{equation}
\par
Particularly, in the orthogonal case, since
$$\bfg_0(i)=\bfh_0(i)=\sqrt{\bff(i)},~~~~\bff(N+1-i)=2-\bff(i),~~~~i=1,..., N,$$ 
we have
$$c_i(\bfh_0):=\sqrt{|\bfh_0(i)|^2+|\bfh_0(N+1-i)|^2}=\sqrt{2},$$
and consequently
\begin{equation}\label{eq:-sigma-A-1}
A_1=\max_{2\le i\le r}\sqrt{\frac{2(2-\bff(i))}{\lambda_i}},~~~~
A_2=\max_{s+1\le i\le N}\sqrt{\frac{2(2-\bff(i))}{\lambda_i}}.
\end{equation}
\par
The discussion proves  the following theorem on the approximation error
of the lowpass channel.
\begin{thm}
If $\bfg_0(N)=0$, then
$$\|\big(\bfI_N-\bfF_{g_0}\bfB_L\bfA_L\bfF_{h_0}\big)\bfx\|_2
\le \frac{1}{2}\big(A_1\sigma_1(\bfx)^{1/2}+ A_2\sigma_2(\bfx)^{1/2}\big),$$
where, $\sigma_1(\bfx),~\sigma_2(\bfx)$ and $A_1, A_2$ are defined by 
\eqref{eq:-sigma-A}. In the orthogonal case, $A_1, A_2$ is also defined by
\eqref{eq:-sigma-A-1}.
\end{thm}
\par
By \eqref{eq:smooth-in-freq} we have that 
$\sigma_1(\bfx)+\sigma_2(\bfx)\le S_2(\bfx)$ and consequently
$$\|\big(\bfI_N-\bfF_{g_0}\bfB_L\bfA_L\bfF_{h_0}\big)\bfx\|_2
\le \frac{1}{2}\sqrt{A_1^2+A_2^2}\sqrt{S_2(\bfx)},$$
which implies that  the smoother the signal the smaller the approximation error.
Particularly, if $\bfx$ is in the Paley–Wiener space 
$\mathrm{PW}_r(\mathcal{G)}:=\{\bfx|\hat{\bfx}(k)=0,~k>r\}$
\cite{huang2021approximation},
then $\sigma_2(\bfx)=0$ and $\|\big(\bfI_N-\bfF_{g_0}\bfB_L\bfA_L\bfF_{h_0}\big)\bfx\|_2\le\frac{1}{2}A_1\sqrt{\sigma_1(\bfx)}$. In this case the lowpass channel of
idealFB gives the best approximation error $0$ since $A_1=0$.

\subsubsection{Approximation Error: Experiments}
\begin{figure}[hbtp]
\begin{center}
\includegraphics[scale=0.65]{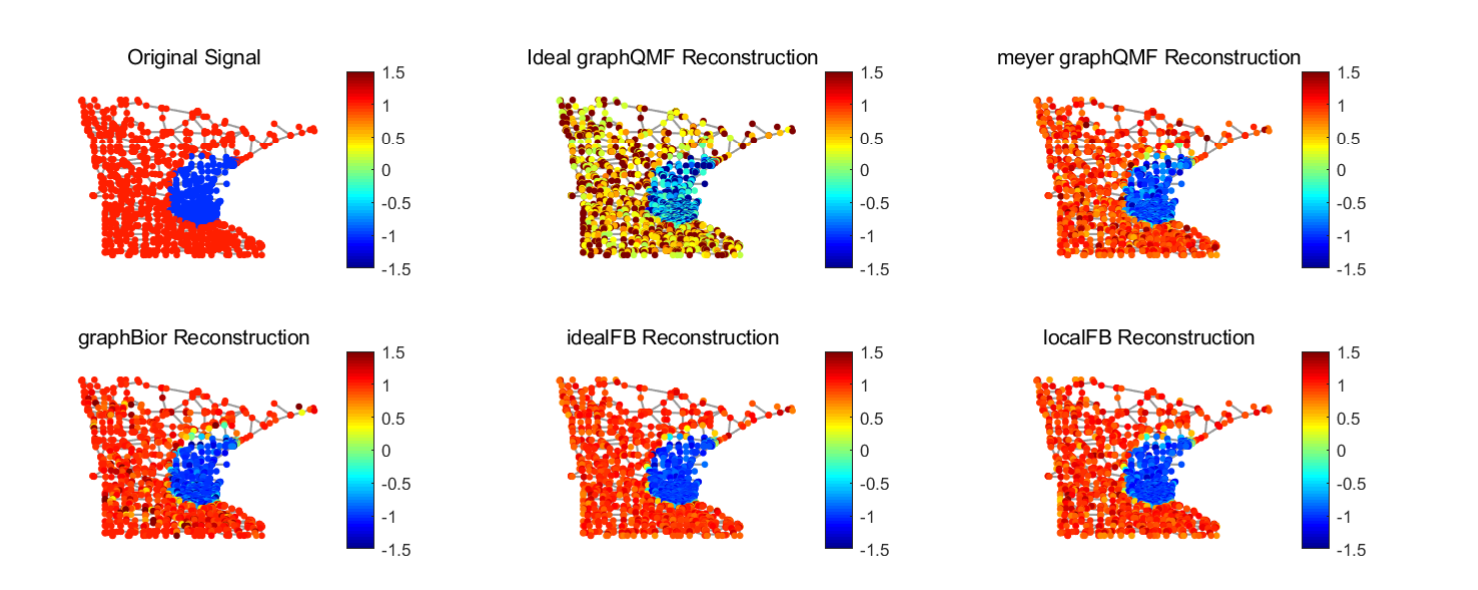}
\end{center}
\caption{\footnotesize Reconstructed signal using only lowpass wavelet coefficients
by five different methods: graphQMF-ideal (top-middle), graphQMF-meyer (top-right),
graphBior (bottom-left), idealFB (bottom-middle) and localFB (bottom-right). 
The top-left is the original signal.
}
\label{minnesota}
\end{figure}

In this section, experiments are implemented to check the 
the approximation error of the lowpass channels of the two types of filter banks: 
localFB and idealFB. 
The experimental results are also compared with graphQMF-meyer and
graphQMF-ideal proposed by Narang and Ortega in \cite{narang2012perfect}
 and graphBior in their follow-up work \cite{narang2013compact}. 
We conduct experiments on the Minnesota traffic graph with normalized Laplacian, where the graph signal is the one used in \cite{narang2012perfect}. All the results are shown
in Figure \ref{minnesota}: The top left is the original signal, with the color of vertices representing the signal values. One layer of decomposition is conducted, and only the lowpass-channel output is going to be used for reconstruction. Since 
the Minnesota traffic graph is not bipartite, the methods proposed in \cite{narang2012perfect} and \cite{narang2013compact} need to decompose 
it into two bipartite subgraphs. Then a two-dimensional filter bank implementation is performed on these subgraphs, producing four channels: LL, LH, HL and HH channel, each channel contains $1188$, $404$, $0$ and $1050$ samples respectively. The reconstruction is done by the wavelet coefficients from LL channel only. While for methods proposed in this paper, we only have two channels: L and H channel. We only use the wavelet coefficients from L channel, containing $1321$ samples, for reconstruction. We calculate the 
SNR and relative error (denoted as RE) of each reconstructed signal using only lowpass wavelet coefficients (denoted as reconSNR and reconRE), which are defined by
\begin{align*}
\rm SNR:= 10\log_{10}(\frac{\|\bff\|_2^2}{\|\bff-\bff_r\|^2_2}),~~~~
\rm RE:=\frac{\|\bff-\bff_r\|_2}{\|\bff\|_2},
\end{align*}
where $\bff$ is the original signal and $\bff_r$ is the reconstructed signal. 
The REs of the overall reconstruction (perfect reconstruction) of each model are also 
calculated (denoted as totalRE). For experiments of graphQMF and gaphBior, we use the codes provided by Narang and Ortega, see  \textit{Biorth\_filterbank\_demo2} and \textit{QMF\_filterbank\_demo\_2} in 
\textcolor{blue}{\href{http://biron.usc.edu/wiki/index.php/Graph_Filterbanks}{Graph Filterbanks}}. The graphQMFs used in the experiments are based on the $24$-th order polynomial approximation of meyer kernel and ideal kernel respectively. The degrees of lowpass and highpass kernels of graphBior are 16 and 17. 
\par
All the experimental results are listed in Table \ref{table:RE}.
It is easy to see that the totalRE of the two graphQMFs are much larger than the other $3$ methods because they use a polynomial approximation of the kernels instead of the exact kernels. In particular, graphQMF based on the ideal kernel performs the worst because it can not be well approximated by a low-order polynomial. From Table \ref{table:RE}, the methods proposed in this paper have perfect reconstruction and perform better in the approximation.
The reconstruction signals are displayed in Figure \ref{minnesota}.  
\begin{table}[htp]
\caption{SNR and relative error of five different methods}
\begin{center}
\begin{tabular}{|c|c|c|c|c|c|}
\hline
&    graphQMF-meyer &   graphQMF-ideal   & graphBior & idealFB & localFB\\
\hline
reconSNR & 12.5647  &  0.9000  &     11.8843  &     15.6612  &     15.0422\\
\hline
reconRE &  0.2354  &  0.9016  &      0.2546  &      0.1648  &      0.1770\\
\hline
totalRE   &  0.0030  &  0.8406  &  9.6305e-06  &  5.2826e-15  &  5.4851e-15\\
\hline
\end{tabular}
\end{center}
\label{table:RE}
\end{table}

%

\bibliographystyle{plain}

\end{document}